\documentclass{IEEEtran}
\usepackage{cite}
\usepackage{amsmath,amssymb,amsfonts}
\usepackage{algorithmic}
\usepackage{graphicx}
\usepackage{textcomp}
\def\BibTeX{{\rm B\kern-.05em{\sc i\kern-.025em b}\kern-.08em
    T\kern-.1667em\lower.7ex\hbox{E}\kern-.125emX}}

\usepackage{balance}
\usepackage{amssymb}
\usepackage{afterpage}
\usepackage{booktabs}
\usepackage{tabularx}
\usepackage{adjustbox}
\usepackage{gensymb}
\usepackage{multirow}
\usepackage{bm} 
\usepackage[numbers]{natbib}
\usepackage{xcolor}
\usepackage[caption=false,font=normalsize]{subfig}

\PassOptionsToPackage{draft}{hyperref} 
\usepackage{pdfcomment}
\newcommand{\beqref}[1]{\eqref{#1}}

\usepackage{hyperref}
\usepackage{pdfcomment}

\begin{document}
\title{Measurement-Based Ultra-Massive MIMO Statistical Channel Characterization and System Performance Evaluation for\\ UMi Environments at 15 GHz FR3 Spectrum}
\author{Panpan Shi, \IEEEmembership{Student Member, IEEE}, Yang Wang, \IEEEmembership{Senior Member, IEEE}, Xi Liao, \IEEEmembership{Senior Member, IEEE}, \\Tianhao Li, \IEEEmembership{Student Member, IEEE}, Jiliang Zhang, \IEEEmembership{Senior Member, IEEE} and Jie Zhang, \IEEEmembership{Senior Member, IEEE}

\thanks{This work has been submitted to the IEEE for possible publication. Copyright may be transferred without notice, after which this version may no longer be accessible. 
\par This work was supported by the Natural Science Foundation of Chongqing under Grant CSTB2025YITP-QCRCX0045. (Corresponding author: Xi Liao.)
\par Panpan Shi, Yang Wang, Xi Liao and Tianhao Li are with the School of Communications and Information Engineering, Chongqing University of Posts and Telecommunications, Chongqing 400065, China (e-mail: d240101024@stu.cqupt.edu.cn; wangyang@cqupt.edu.cn; liaoxi@cqupt.edu.cn; d240101012@stu.cqupt.edu.cn).

\par Jiliang Zhang is with the College of Information Science and Engineering, Northeastern University, Shenyang 110819, China (e-mail: zhangjiliang1@mail.neu.edu.cn).

\par Jie Zhang is with the Department of Electronic and Electrical Engineering, The University of Sheffield, S10 2TN Sheffield, U.K (e-mail: jie.zhang@sheffied.ac.uk).

}}

\maketitle

\begin{abstract}
This paper presents a detailed measurement campaign and a comprehensive analysis of 15 GHz ultra-massive multiple-input multiple-output (UM-MIMO) channels tailored for the urban microcell (UMi) environment. Channel sounding is performed over 14.875--15.125 GHz using a time-domain platform comprising a 128-element L-shaped transmit array and a 64-element square receive array. Four representative scenarios are investigated, namely near-field line-of-sight (LoS), near-field foliage-shaded, far-field foliage-shaded, and far-field LoS street canyon scenarios, resulting in 81 distinct transmit-receive links. Based on the measured data, conventional channel characteristics, including path loss, power delay angle profiles, delay spread, and angular spread, are characterized, while UM-MIMO-specific phenomena associated with near-field effects, spatial non-stationarity (SNS), and channel hardening (CHD) are quantitatively analyzed. Channel capacity is further evaluated to reveal the effects of different UMi propagation conditions on system performance. The reported results provide empirical support for the new mid-band spectrum (6--24 GHz, including Frequency Range 3 (FR3)) UM-MIMO channel modeling and offer practical guidance for the design and deployment of future sixth-generation (6G) microcell networks.

\end{abstract}


\section{Introduction}\label{sec:introduction}
\IEEEPARstart{F}{uture} sixth-generation (6G) wireless systems are expected to support ubiquitous coverage, extremely high data rates, and seamless connectivity for a vast number of devices and services. Meeting these requirements will necessitate both new spectrum resources and more advanced multi-antenna transmission technologies \cite{6G1}. Against this research backdrop, the 6--24 GHz, known as frequency range 3 (FR3) band, has become a key spectrum candidate for beyond-5G (B5G) and future 6G networks. For example, the 2023 World Radiocommunication Conference (WRC-23) explicitly stated that future WRC cycles would prioritize the following frequency bands for international mobile communications: 4.4--4.8 GHz, 7.125--8.4 GHz, and 14.8--15.35 GHz. In December 2023, the 3rd Generation Partnership Project (3GPP) Technical Specification Group Radio Access Network Release 19 initiated research work within the 7--24 GHz FR3 band \cite{NMB3}. At the same time, ultra-massive multiple-input multiple-output (UM-MIMO) technology can exploit the abundant spatial degrees of freedom available at base stations, thereby enhancing multiplexing capability, system capacity, and spectral efficiency \cite{UMMIMO1}, \cite{UMMIMO2}. Consequently, the combination of FR3 spectrum and UM-MIMO technology is expected to become a key enabling technology for future 6G networks \cite{Combination1}, \cite{Combination2}.

\par Accurate channel models are crucial to the design, optimization and performance evaluation of UM-MIMO systems in the FR3 band. Channel measurements provide real-world channel data and form a vital foundation for establishing reliable channel models. Measurement-based channel research typically encompasses channel measurement, parameter extraction, statistical characteristic analysis and channel modeling \cite{XLMIMOprerequisite}. UM-MIMO channel sounders are key equipment for conducting channel measurements. The existing FR3-band UM-MIMO channel sounders can be classified from two perspectives. In terms of sounding principle, channel sounders are typically categorized as frequency-domain sounders based on vector network analysers (VNAs) and time-domain sounders based on pseudo-random (PN) sequence correlation \cite{Channelsounderscategories2}. In terms of array implementation, these mainly include schemes based on virtual antenna arrays (VAAs) \cite{NMB3} and those based on real antenna arrays (RAAs) \cite{RAA128x8PN}. The former synthesizes a large-aperture array through spatial scanning using a small number of physical antennas, whereas the latter utilizes multiple physical elements to acquire the channel directly.
\begin{table*}[t]
	\centering
	\setlength{\tabcolsep}{2pt} 
	\caption{Summary of FR3 Bands UM-MIMO Channel Measurements and Characterisation}
	\label{tab:Summary}
	\resizebox{\textwidth}{!}{ 
		\begin{tabular}{      
				>{\centering\arraybackslash} m{0.7cm}   
				>{\centering\arraybackslash} m{1.9cm}   
				>{\centering\arraybackslash} m{1.5cm}   
				>{\centering\arraybackslash} m{1.5cm}   
				>{\centering\arraybackslash} m{1.5cm}   
				>{\centering\arraybackslash} m{1.2cm}   
				>{\centering\arraybackslash} m{1.5cm}   
				>{\centering\arraybackslash} m{1.3cm}   
				>{\centering\arraybackslash} m{1.3cm}   
				>{\centering\arraybackslash} m{0.8cm}   
				>{\centering\arraybackslash} m{0.6cm}   
				>{\centering\arraybackslash} m{1cm}     
			}
			\toprule
			\multirow{4}{*}{\begin{tabular}{c}Ref.\end{tabular}}  &
			\multirow{4}{*}{\begin{tabular}{c}Sounder    \\implementation\\ principle\end{tabular}} &
			\multirow{4}{*}{\begin{tabular}{c}Center\\frequency     \\ (Bandwidth)\end{tabular}} &
			\multirow{4}{*}{\begin{tabular}{c}Antenna    \\array         \\ type\end{tabular}} &
			\multirow{4}{*}{\begin{tabular}{c}Antenna    \\array 		 \\ configuration\end{tabular}} &
			\multirow{4}{*}{\begin{tabular}{c}Scenarios \end{tabular}}  &
			\multirow{4}{*}{\begin{tabular}{c}Measurement\\distance 	 \\ range\end{tabular}} &
			\multicolumn{2}{c}{TCC} & \multicolumn{3}{c}{NCC} \\
			\cmidrule(lr){8-9} \cmidrule(lr){10-12} & & & & & & & 
			
			\multirow{2}{*}{\begin{tabular}{c}Large-scale \\ fading\end{tabular}} & 
			\multirow{2}{*}{\begin{tabular}{c}Small-scale \\ fading\end{tabular}} & 
			\multirow{2}{*}{\begin{tabular}{c}NF \\ effects\end{tabular}} & 
			\multirow{2}{*}{\begin{tabular}{c}{SNS}\end{tabular}} &
			\multirow{2}{*}{\begin{tabular}{c}CHD \end{tabular}} \\	[3.5 mm]
			\midrule	
			
			\multirow{2}{*}{\centering [10]} & \multirow{2}{*}{\centering Frequency} & \multirow{2}{*}{\begin{tabular}{c}10.1 GHz\\ (500 MHz)\end{tabular}} &
			\multirow{2}{*}{\begin{tabular}{c}VAA,\\ UPA\end{tabular}}  & \multirow{2}{*}{100 $\times$ 25} 	&
			\multirow{2}{*}{\begin{tabular}{c}UMi \end{tabular}} & \multirow{2}{*}{\centering 16--36.5 m} & \multirow{2}{*}{\centering $\mathbf{\checkmark}$} & \multirow{2}{*}{\centering $\mathbf{\checkmark}$} & \multirow{2}{*}{\centering $\mathbf{\times}$} & 	
			\multirow{2}{*}{\centering $\mathbf{\times}$} & \multirow{2}{*}{\centering $\mathbf{\times}$}  	\\[4 mm]				
			
			\multirow{2}{*}{\centering [11]} & \multirow{2}{*}{\centering Frequency} & \multirow{2}{*}{\begin{tabular}{c}15 GHz\\ (4 GHz)\end{tabular}} &
			\multirow{2}{*}{\begin{tabular}{c}VAA,\\ UPA\end{tabular}}  & \multirow{2}{*}{40 $\times$ 40} 	&
			\multirow{2}{*}{\begin{tabular}{c}Building \\ roof\end{tabular}} & \multirow{2}{*}{\centering 20 m} & \multirow{2}{*}{\centering ${\checkmark}$} &
			\multirow{2}{*}{\centering $\mathbf{\checkmark}$} & \multirow{2}{*}{\centering $\mathbf{\times}$} & 	
			\multirow{2}{*}{\centering $\mathbf{\times}$} & \multirow{2}{*}{\centering $\mathbf{\times}$}  	\\[4 mm]
							
			\multirow{2}{*}{\centering [12]} & \multirow{2}{*}{\centering Frequency} & \multirow{2}{*}{\begin{tabular}{c}15 GHz\\ (2 GHz)\end{tabular}} &
			\multirow{2}{*}{\begin{tabular}{c}VAA,\\ UCA\end{tabular}}  & \multirow{2}{*}{720 $\times$ 1} 	&
			\multirow{2}{*}{\begin{tabular}{c}Indoor \\ corridor\end{tabular}} & \multirow{2}{*}{\centering 5 m} & \multirow{2}{*}{\centering ${\times}$} &
			\multirow{2}{*}{\centering $\mathbf{\times}$} & \multirow{2}{*}{\centering $\mathbf{\checkmark}$} & 	
			\multirow{2}{*}{\centering $\mathbf{\times}$} & \multirow{2}{*}{\centering $\mathbf{\times}$}  	\\[4 mm]	
			
			\multirow{2}{*}{\centering [13]} & \multirow{2}{*}{\centering Frequency} & \multirow{2}{*}{\begin{tabular}{c}11/16 GHz\\ (2 GHz)\end{tabular}} &
			\multirow{2}{*}{\begin{tabular}{c}VAA,\\ URA\end{tabular}}  & \multirow{2}{*}{\begin{tabular}{c}51 $\times$ 51,\\76 $\times$ 76\end{tabular}} &
			\multirow{2}{*}{\begin{tabular}{c}Indoor \\ office\end{tabular}} & \multirow{2}{*}{\centering 1.7--2.7 m} & \multirow{2}{*}{\centering ${\checkmark}$} &
			\multirow{2}{*}{\centering $\mathbf{\checkmark}$} & \multirow{2}{*}{\centering $\mathbf{\times}$} & 	
			\multirow{2}{*}{\centering $\mathbf{\checkmark}$} & \multirow{2}{*}{\centering $\mathbf{\times}$}  	\\[4 mm]	
			
			\multirow{2}{*}{\centering [14]} & \multirow{2}{*}{\centering Time} & \multirow{2}{*}{\begin{tabular}{c}6 GHz\\ (200 MHz)\end{tabular}} &
			\multirow{2}{*}{\begin{tabular}{c}VAA,\\ UPA\end{tabular}}  & \multirow{2}{*}{16 $\times$ 256} 	&
			\multirow{2}{*}{\begin{tabular}{c}UMi/UMa \end{tabular}} & \multirow{2}{*}{\centering -} & \multirow{2}{*}{\centering $\mathbf{\times}$} & \multirow{2}{*}{\centering $\mathbf{\checkmark}$} & \multirow{2}{*}{\centering $\mathbf{\times}$} & 	
			\multirow{2}{*}{\centering $\mathbf{\times}$} & \multirow{2}{*}{\centering $\mathbf{\times}$}  	\\[4 mm]	
			
			\multirow{2}{*}{\centering [15]} & \multirow{2}{*}{\centering Time} & \multirow{2}{*}{\begin{tabular}{c}13 GHz\\ (400 MHz)\end{tabular}} &
			\multirow{2}{*}{\begin{tabular}{c}VAA,\\ UPA\end{tabular}}  & \multirow{2}{*}{\begin{tabular}{c}64 $\times$ 128 \end{tabular}} &
			\multirow{2}{*}{\begin{tabular}{c}Indoor \\ office\end{tabular}} & \multirow{2}{*}{\centering 3--12 m} & \multirow{2}{*}{\centering ${\times}$} &
			\multirow{2}{*}{\centering $\mathbf{\times}$} & \multirow{2}{*}{\centering $\mathbf{\times}$} & 	
			\multirow{2}{*}{\centering $\mathbf{\checkmark}$} & \multirow{2}{*}{\centering $\mathbf{\times}$}  	\\[4 mm]		
			
			\multirow{2}{*}{\centering [16]} & \multirow{2}{*}{\centering Time} & \multirow{2}{*}{\begin{tabular}{c}13 GHz\\ (400 MHz)\end{tabular}} &
			\multirow{2}{*}{\begin{tabular}{c}VAA,\\ UPA\end{tabular}}  & \multirow{2}{*}{\begin{tabular}{c}64 $\times$ 128 \end{tabular}} &
			\multirow{2}{*}{\begin{tabular}{c}UMa\end{tabular}} & \multirow{2}{*}{\centering 50--200 m} & \multirow{2}{*}{\centering ${\times}$} &
			\multirow{2}{*}{\centering $\mathbf{\times}$} & \multirow{2}{*}{\centering $\mathbf{\checkmark}$} & 	
			\multirow{2}{*}{\centering $\mathbf{\checkmark}$} & \multirow{2}{*}{\centering $\mathbf{\times}$}  	\\[4 mm]
			
			\multirow{2}{*}{\centering [17]} & \multirow{2}{*}{\centering Time} & \multirow{2}{*}{\begin{tabular}{c}11 GHz\\ (400 MHz)\end{tabular}} &
			\multirow{2}{*}{\begin{tabular}{c}RAA,\\ UCA\end{tabular}}  & \multirow{2}{*}{\begin{tabular}{c}24 $\times$ 24 \end{tabular}} &
			\multirow{2}{*}{\begin{tabular}{c}Indoor\\corridor\end{tabular}} & \multirow{2}{*}{\centering 3--12 m} & \multirow{2}{*}{\centering ${\checkmark}$} &
			\multirow{2}{*}{\centering $\mathbf{\checkmark}$} & \multirow{2}{*}{\centering $\mathbf{\times}$} & 	
			\multirow{2}{*}{\centering $\mathbf{\times}$} & \multirow{2}{*}{\centering $\mathbf{\times}$}  	\\[4 mm]
			
			\multirow{3}{*}{\centering This work} & \multirow{3}{*}{\centering Time} & \multirow{3}{*}{\begin{tabular}{c}15 GHz\\ (250 MHz)\end{tabular}} &
			\multirow{3}{*}{\begin{tabular}{c}RAA,\\ L-shaped\\ ULA\end{tabular}}  & \multirow{3}{*}{\begin{tabular}{c}64 $\times$ 128 \end{tabular}} &
			\multirow{3}{*}{\begin{tabular}{c}UMi\end{tabular}} & \multirow{3}{*}{\centering 21--222 m} & \multirow{3}{*}{\centering ${\checkmark}$} &
			\multirow{3}{*}{\centering $\mathbf{\checkmark}$} & \multirow{3}{*}{\centering $\mathbf{\checkmark}$} & 	
			\multirow{3}{*}{\centering $\mathbf{\checkmark}$} & \multirow{3}{*}{\centering $\mathbf{\checkmark}$}  	\\[6 mm]
			\bottomrule
		\end{tabular}
	}
	\vspace{-0.5em}
\end{table*}

\par Because commercial VNA platforms are capable of supporting channel measurements in the FR3 band, the combination of a VNA and a VAA has been widely employed as a channel sounder in previous studies. For instance, reported measurements have covered 9.85--10.35 GHz \cite{VAA100x2510GHzVNA}, 13--17 GHz \cite{VAA40x4013-17GHzVNA}, 14--16 GHz \cite{VAAUCA15GHzVNA}, and 10--13 GHz/15--17 GHz \cite{VAAMulti-frequencyVNA}. The array types employed include uniform planar array (UPA) \cite{VAA100x2510GHzVNA}, \cite{VAA40x4013-17GHzVNA}, uniform circular array (UCA) \cite{VAAUCA15GHzVNA} and uniform rectangular array (URA) \cite{VAAMulti-frequencyVNA}. The scenarios measured include UMi \cite{VAA100x2510GHzVNA}, building rooftops \cite{VAA40x4013-17GHzVNA}, indoor corridors \cite{VAAUCA15GHzVNA} and offices \cite{VAAMulti-frequencyVNA}. To further satisfy the requirements of long-range measurements, VAA-based time-domain UM-MIMO channel measurement sounders have been employed. For example, Miao et al. established a UPA-based channel measurement platform with a frequency range of 3--16 GHz and a maximum measurement bandwidth of 2 GHz \cite{NMB3}. Channel measurements were conducted in the 5.9--6.1 GHz \cite{VAA16x2566GHzPN3} and 12.8--13.2 GHz \cite{VAA64x12813GHzPN1}, \cite{VAA64x12813GHzPN2} bands, with measurement scenarios including UMi \cite{VAA16x2566GHzPN3}, UMa \cite{VAA16x2566GHzPN3}, \cite{VAA64x12813GHzPN2} and indoor office \cite{VAA64x12813GHzPN1} environments. Channel measurement campaigns based on the VAA architecture provide valuable insights into the channel propagation characteristics of UM-MIMO in the FR3 band. 

\par However, the process of forming a large-scale array using the VAA approach typically requires mechanical scanning, which may render the sounder unsuitable for time-varying channels. Furthermore, VAA-based UM-MIMO channel sounding systems generally do not account for mutual coupling between elements in real arrays. To meet the demands of long-range dynamic channel measurements and more realistic system evaluation, time-domain channel sounders based on RAAs are attracting increasing attention. For example, Kim et al. constructed a 24 $\times$ 24 MIMO channel sounder using a dual-polarised 12-element UCA \cite{RAA24x2411GHzPN}. Achieving high-resolution angle estimation in both azimuth and elevation typically requires a full UPA, which entails high hardware complexity and cost. To address this issue, this paper adopts an L-shaped array, whose two orthogonal linear subarrays independently estimate azimuth and elevation, enabling two-dimensional angular resolution with significantly reduced hardware requirements \cite{Larray}.

\par Table \ref{tab:Summary} summarizes representative UM-MIMO measurement campaigns in the FR3 band, covering sounder implementation principles, bandwidth, array configurations, scenarios, measurement distance ranges, and the investigated channel characteristics. It is clear that most existing measurement campaigns focus primarily on indoor environments, lower-frequency UMi environments and urban macrocell (UMa) environments. Furthermore, the reported analyses primarily concern traditional channel characteristics (TCC) such as large-scale and small-scale fading. By contrast, comprehensive measurement campaigns and channel characterization of 15 GHz UM-MIMO propagation in UMi environments remain limited, particularly with regard to real-world antenna array configurations and new channel characteristics (NCC) such as near-field (NF) effects, spatial non-stationarity (SNS) and channel hardening (CHD). The main contributions of this paper are summarized as follows.
\begin{itemize}
	\item{UM-MIMO channel measurement campaigns with a centre frequency of 15 GHz, a bandwidth of 250 MHz, and a real-array size of 64 $\times$ 128 are carried out in near-field LoS, near-field foliage-shaded, far-field foliage-shaded and far-field LoS street canyon scenarios in the UMi environment. In addition, the UM-MIMO measured dataset is constructed for the UMi environment, comprising 81 distinct transmit-receive links.}
	
	\item{The traditional and new channel statistical characteristics of 15 GHz UM-MIMO near-field and far-field channels are investigated and analyzed. The investigated TCC include path loss, power delay angle profile, delay spread, and angle spread, while the investigated NCC include near-field effects, SNS and CHD. This research will lay an important foundation for future studies of UM-MIMO channel models in the FR3 band.}
		
	\item{The performance of 15 GHz UM-MIMO systems is evaluated in terms of channel capacity. By comparing the average channel capacity in far-field and near-field scenarios with that of an i.i.d. Rayleigh channel, the performance of 15 GHz UM-MIMO systems is clarified. This provides the guidance for the deployment and optimization of 15 GHz UM-MIMO wireless communication networks.}
	
\end{itemize}
\begin{figure}[t]
	\centering
	\subfloat[]{%
		\includegraphics[width=1\columnwidth]{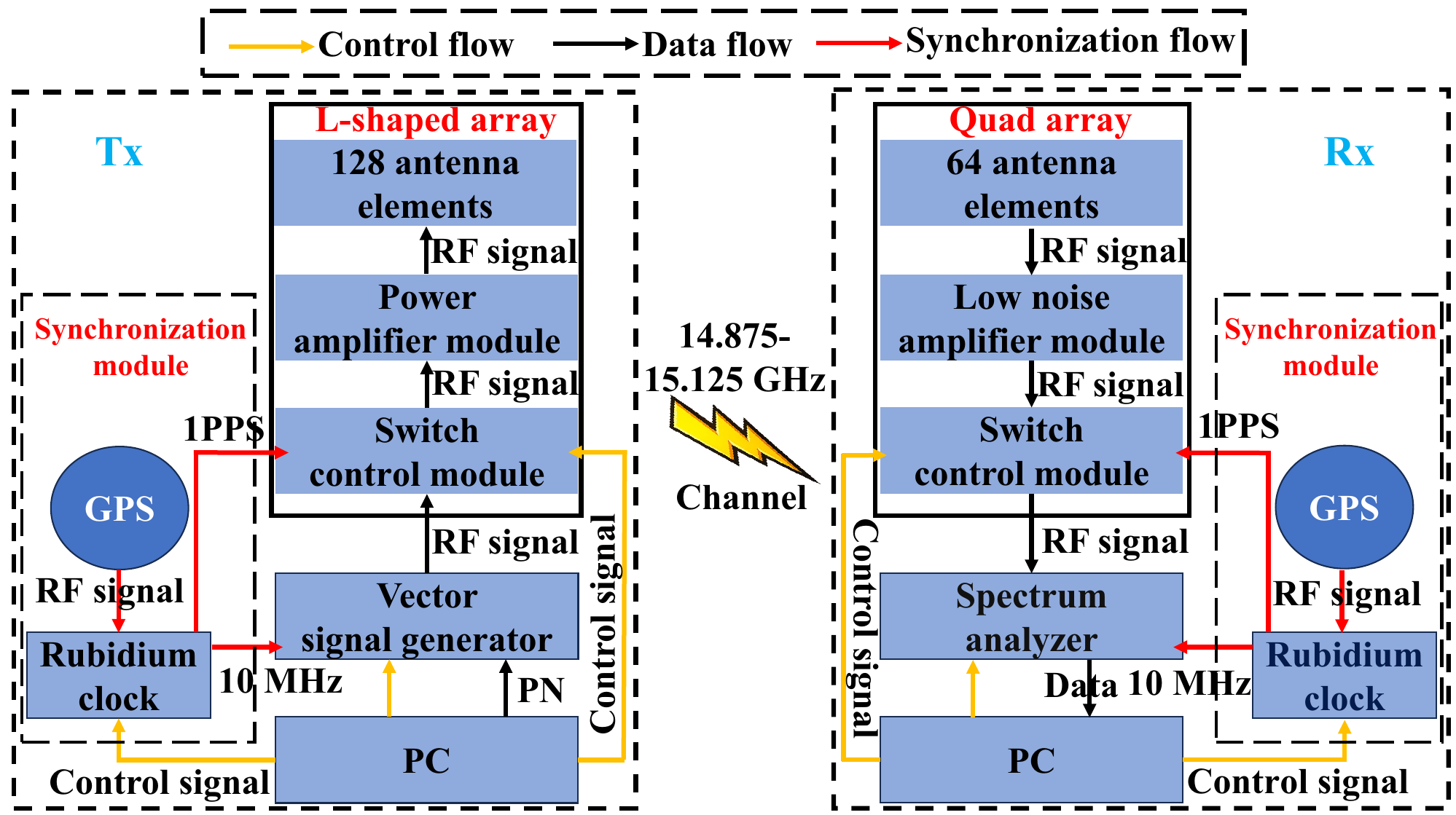}%
		\label{fig:channel sounder system}
	}
	\hfill
	\subfloat[]{%
		\includegraphics[width=0.49\columnwidth]{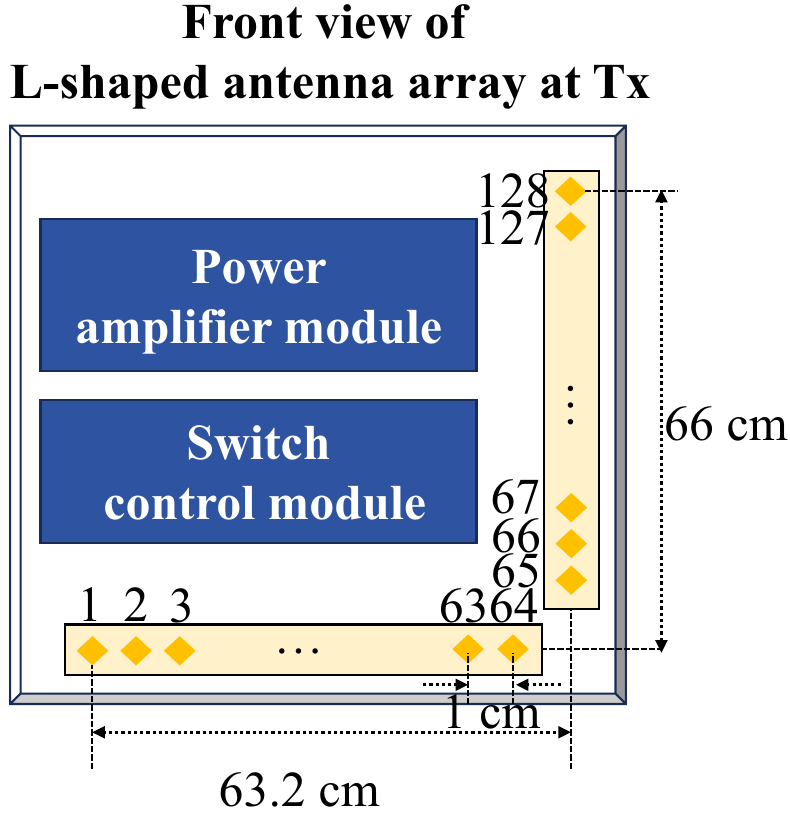}%
		\label{fig:Tx array}%
	}
	\subfloat[]{%
		\includegraphics[width=0.49\columnwidth]{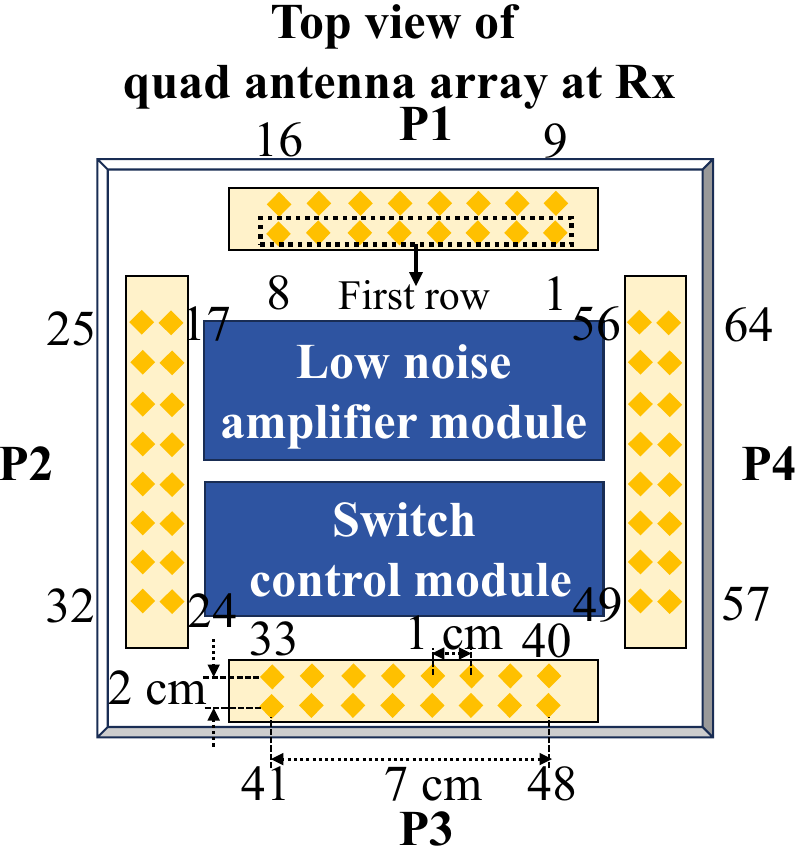}%
		\label{fig:Rx array}%
	}
	\caption{The schematic diagram of UM-MIMO channel sounder system, (a) 14.875-15.125 GHz channel sounder, (b) Tx L-shaped antenna array, and (c) Rx quad antenna array.}
	\label{fig:sounder}
\end{figure}

\par The remainder of this paper is organized as follows. In Sec. \ref{sec:Measurement}, we describe the details of the 15 GHz UM-MIMO channel measurement platform as well as the channel measurement campaigns in the four communication scenarios. The traditional channel characteristics such as large-scale path modeling, delay spread, and angle spread in four communication scenarios are investigated in Sec. \ref{sec:Traditional}. Then, the new channel characteristics such as near-field effects, spatial non-stationarity and channel hardening of UM-MIMO channels are investigated in four different communication scenarios in Sec. \ref{sec:new}. The performance of 15 GHz UM-MIMO communication system is evaluated in terms of channel capacity in Sec. \ref{sec:Systems}.  Finally, the paper is concluded in Sec. \ref{sec:Conclusion}.

\begin{table}[t]
	\centering
	\caption{Channel Measurement Configuration}
	\label{tab:measurement_system}
	\renewcommand{\arraystretch}{1.2}
	\begin{tabular}{p{4.5cm} p{1.5cm} p{1.5cm}}
		\toprule
		\text{Parameter} & \text{Symbol} & \text{Value} \\
		\midrule
		Start frequency & $f_{\rm{start}}$ & 14.875 GHz \\
		End frequency & $f_{\rm{end}}$ & 15.125 GHz \\
		Bandwidth & $B_{\rm{sys}}$ & 250 MHz \\
		Time domain resolution & $\Delta \tau$ & 4 ns \\
		Path length resolution & $\Delta d$ & 1.2 m \\
		PN sequence length & ${PN}_{\rm {len}}$ & 1023 \\
		
		Transmit symbol rate of VSG & $R_{\mathrm{VSG}}$ & 250 MSym/s \\
		Output power of VSG & $P_{\mathrm{VSG}}$ & -15 dBm \\
		Number of Tx antenna array elements & $N_{\mathrm{Tx}}$ & 128 \\
		Minimum isolation of Tx switches & ${ISO}_{\mathrm{min,Tx}}$ & 40 dB \\
		Antenna gain at Tx & $G_{\mathrm{Tx}}^{\mathrm{Ant}}$ & 5 dBi \\
		Power amplifier gain at Tx & $G_{\mathrm{Tx}}^{\mathrm{PA}}$ & 30 dB \\
		HPBW of Tx antenna element & ${HPBW}_{\mathrm{Tx}}$ & 120$^\circ$ \\
		Tx antenna polarization & -- & +45$^\circ$ \\
		
		SA sample rate & $R_{\mathrm{SA}}$ & 1 GSa/s \\
		Number of Rx antenna array elements & $N_{\mathrm{Rx}}$ & 64 \\
		Minimum isolation of Rx switches & ${ISO}_{\mathrm{min,Rx}}$ & 40 dB \\
		Antenna gain at Rx & $G_{\mathrm{Rx}}^{\mathrm{Ant}}$ & 5 dBi \\
		LNA gain at Rx & $G_{\mathrm{Rx}}^{\mathrm{LNA}}$ & 35 dB \\
		HPBW of Rx antenna element & ${HPBW}_{\mathrm{Rx}}$ & 120$^\circ$ \\
		Rx antenna polarization & -- & +45$^\circ$ \\
		Average noise floor & $P_{\mathrm{N}}$ & -130 dBm \\
		\bottomrule
	\end{tabular}
\end{table}
\section{Channel Measurement Campaign}\label{sec:Measurement}
This section describes the 15 GHz UM-MIMO channel measurement campaign, including the hardware system specifications, UMi measurement environments, and deployment configurations. The UM-MIMO channel calibration and multipath parameter extraction are also introduced.

\subsection{Channel Measurement System Setup}
\par The 14.875--15.125 GHz UM-MIMO channel measurement system consists of a vector signal generator (VSG), an L-shaped antenna array, a quad antenna array, a spectrum analyzer (SA) and two GPS rubidium clocks, as shown in Fig. \ref{fig:sounder}. In Fig. \ref{fig:sounder}\subref{fig:channel sounder system}, the transmitter (Tx) uses a Rohde \& Schwarz (R\&S) SMW 200A VSG, while a R\&S FSW50 SA is used for signal reception at the receiver (Rx). The UM-MIMO array size is 64 $\times$ 128. Specifically, in Fig. \ref{fig:sounder}\subref{fig:Tx array}, the L-shaped antenna array measures 63.2 cm $\times$ 66 cm and comprises 64 elements in each of the horizontal and vertical directions, a 128-channel switch control module and a power amplifier module.
In Fig. \ref{fig:sounder}\subref{fig:Rx array}, the quad antenna array consists of four 7 cm $\times$ 2 cm subarrays with 16 elements, a 64-channel switch control module and a low-noise amplifier (LNA) module.

\par The measurement system setup is shown in Table \ref{tab:measurement_system}. The sounding bandwidth is ${B_{\rm sys}}=$ 250 MHz, yielding a delay resolution of $\Delta\tau=1/B_{\rm sys}=4$ ns. A PN sequence of length ${PN}_{\rm len}= $ 1023 is used. The effective isotropic radiated power (EIRP) of the Tx array is approximately 15 dBm over 14.875--15.125 GHz. To enhance the measurement dynamic range, the Tx power amplifier gain $G_{\rm{Tx}}^{\rm{PA}}$ and Rx low-noise amplifier gain $G_{\rm{Rx}}^{\rm{LNA}}$ are set to 30 dB and 35 dB, respectively. The measurable path-loss range is determined from the link budget as $PL_{\rm max}={\rm EIRP}+G_{\rm{Rx}}^{\rm{Ant}}-P_{\rm sens}.$ With ${\rm EIRP}\approx$ 15 dBm, $G_{\mathrm{Rx}}^{\mathrm{Ant}}=5$ dBi, and an average noise floor $P_{\mathrm{N}}$ of $\rm {-130}$ dBm, a 20 dB detection margin gives the received power $P_{\rm sens}=-110$ dBm. Hence, $PL_{\rm max}\approx \rm {15}+\rm 5-(-\rm 110)=\rm 130\ {\rm dB}.$ This confirms that the system can measure path loss of at least 130 dB over the 250 MHz bandwidth. To obtain an accurate channel impulse response efficiently, it is necessary to remove the measurement system response through over-the-air (OTA) calibration. The Tx and Rx antenna arrays use identical array elements, and the element polarization is set to +45$^\circ$ during the measurements.

\par During our measurement campaigns, both the Tx and Rx arrays are kept fixed and the Tx antenna array is always oriented toward the first planar (P1) of the Rx array. The three-dimensional (3D) distance between the Tx and Rx is obtained from the GPS latitude and longitude information. To ensure that all Rx points are within the 3 dB beamwidth of the Tx array and thus benefit from the maximum antenna gain, the elevation angle of the Tx array is set to 15$^\circ$. Based on the sizes of the Rx and Tx arrays and the center operating frequency, the near-field boundary is calculated from the Rayleigh distance formula as 42.5 m.

\par In summary, the 14.875--15.125 GHz UM-MIMO broadband channel measurement platform established in this paper features a bandwidth of 250 MHz, a delay resolution of 4 ns, an array size of 64 $\times$ 128, and a maximum measurable path loss of approximately 130 dB. According to the 15 GHz free-space path loss model, the corresponding equivalent maximum measurement distance is approximately 5.0 km, thereby enabling precise measurements of broadband, multipath, high-loss and near-field propagation characteristics under ultra-large-scale array conditions.
\begin{figure}[t]
	\centering
	\includegraphics[width= 3.5 in]{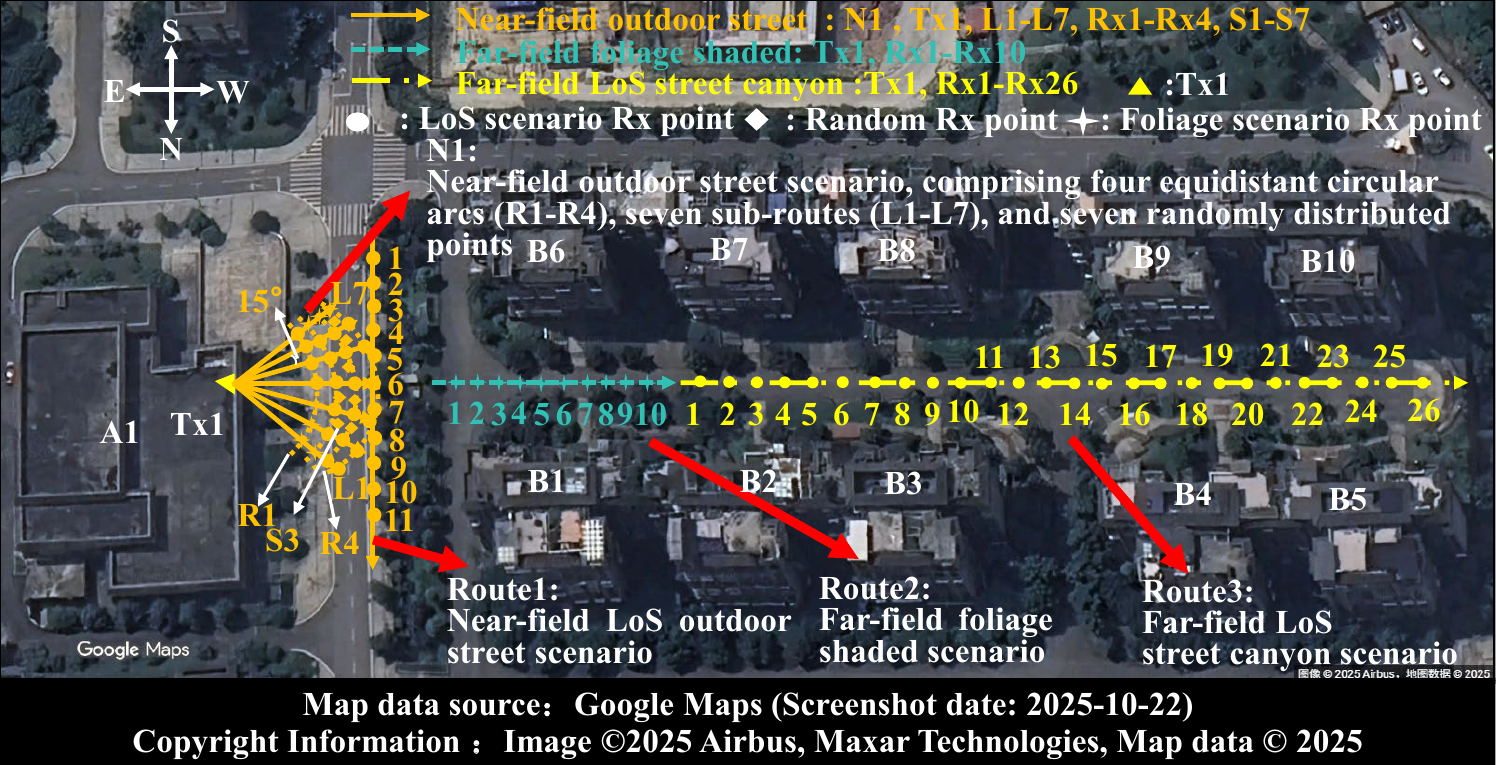}
	\caption{Deployments of the channel measurement campaign in the UMi environment.}
	\label{fig:UMi scenario} 
\end{figure}

\begin{figure}[t]
	\centering
	
	\begin{minipage}[t]{0.49\columnwidth}
		\centering
		\subfloat[]{%
			\includegraphics[width=\linewidth]{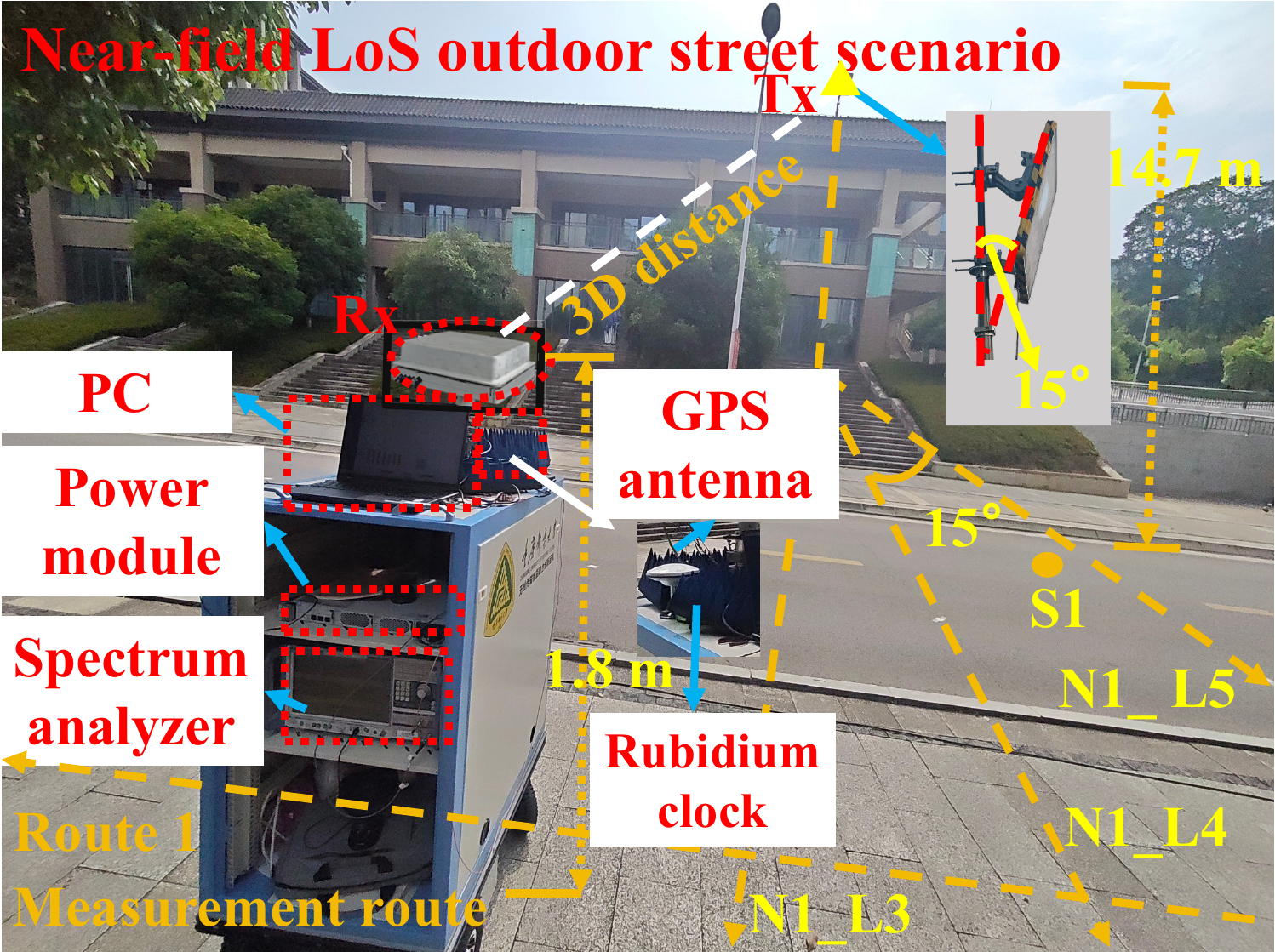}%
			\label{fig:NF LOS}
		}
	\end{minipage}
	\hfill
	\begin{minipage}[t]{0.49\columnwidth}
		\centering
		\subfloat[]{%
			\includegraphics[width=\linewidth]{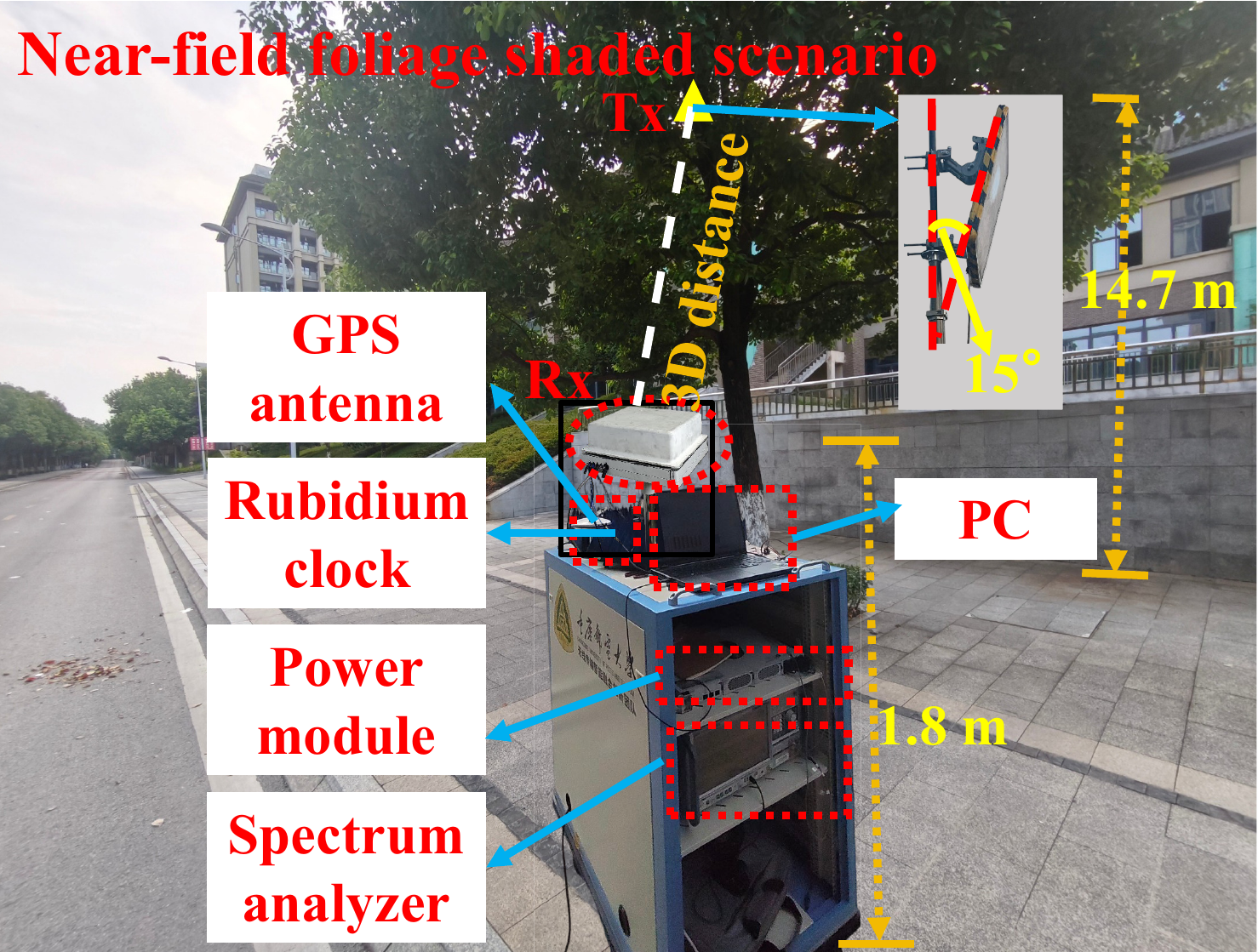}%
			\label{fig:NF Foliage}
		}
	\end{minipage}
	
	\vspace{0.5em}
	
	\begin{minipage}[t]{0.49\columnwidth}
		\centering
		\subfloat[]{%
			\includegraphics[width=\linewidth]{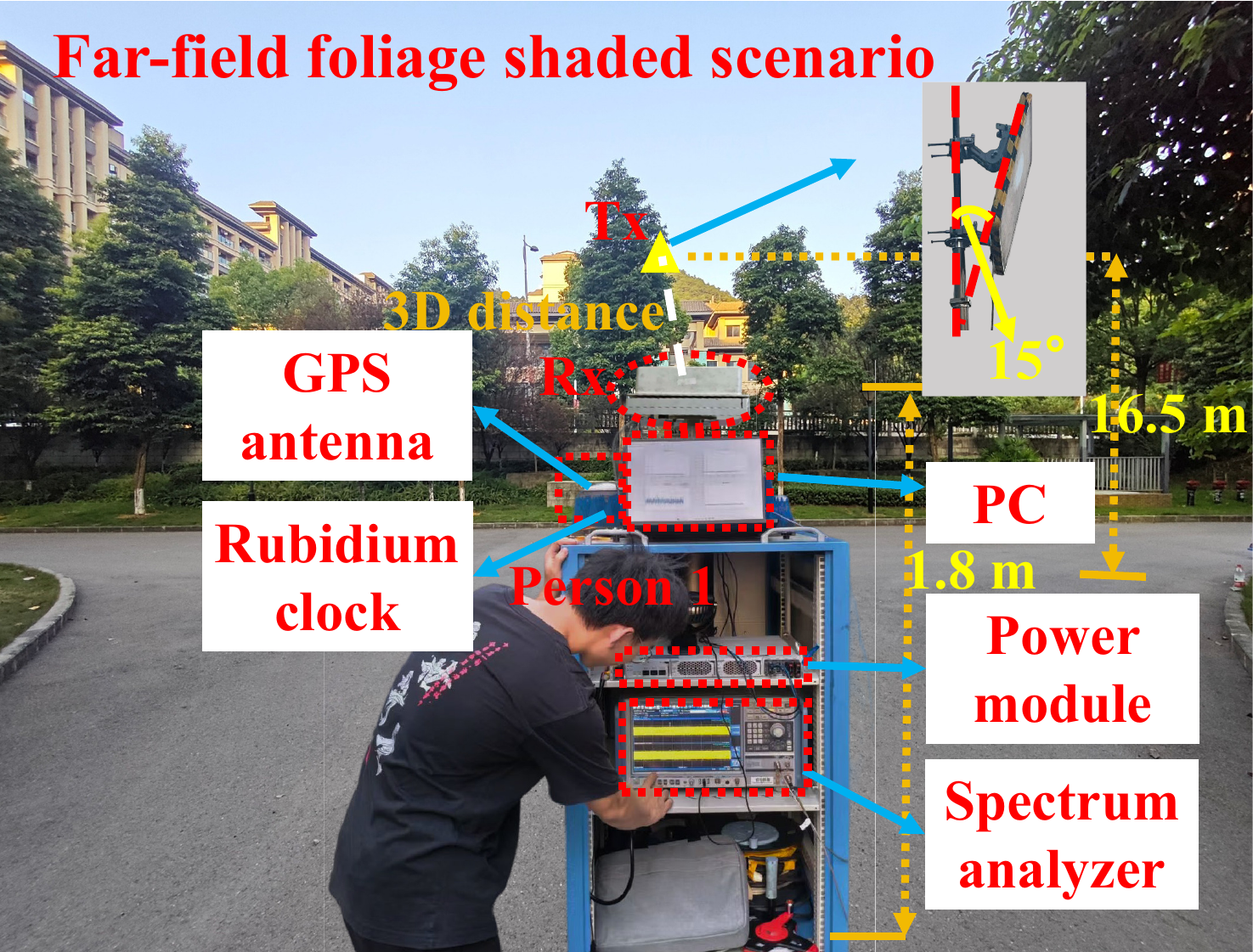}%
			\label{fig:FF Foliage}
		}
	\end{minipage}
	\hfill
	\begin{minipage}[t]{0.49\columnwidth}
		\centering
		\subfloat[]{%
			\includegraphics[width=\linewidth]{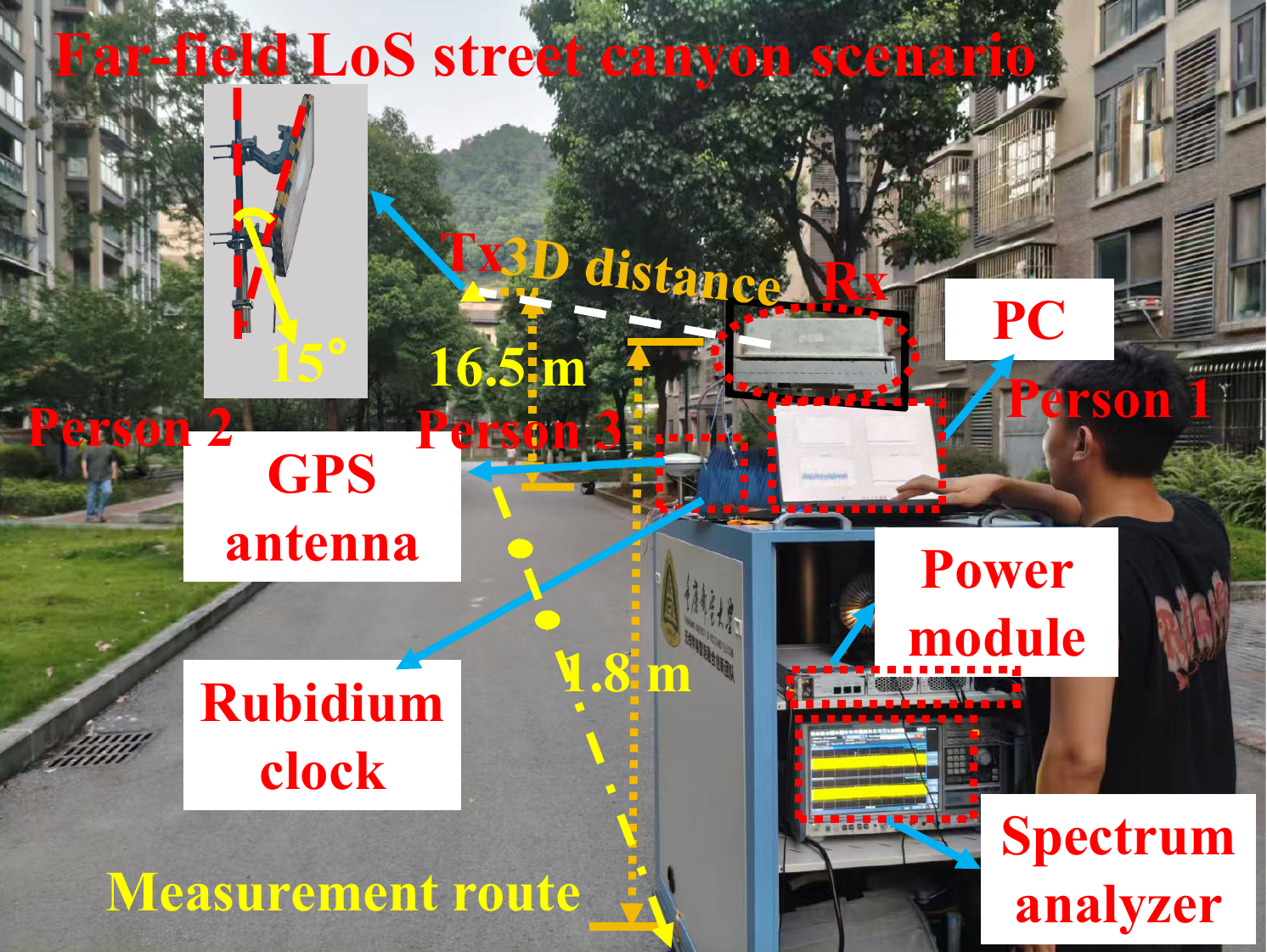}%
			\label{fig:FF LOS}
		}
	\end{minipage}
	
	\caption{Photographs of the UMi environment under (a) near-field outdoor street scenario, (b) near-field foliage shaded scenario, (c) far-field foliage shaded scenario, and (d) far-field LoS street canyon scenario.}
	\label{fig:The photography}
\end{figure}

\begin{table*}[t]
	\centering
	\caption{Channel Measurement Deployments In The UMi Environment}
	\label{tab:measurement deployment}
	
	\renewcommand{\arraystretch}{1}
	\begin{tabular}{
			>{\centering\arraybackslash} m{1.8cm}
			>{\centering\arraybackslash} m{1.2cm}
			>{\centering\arraybackslash} m{1.5cm}
			>{\centering\arraybackslash} m{2.4cm}
			>{\centering\arraybackslash} m{2.1 cm}
			>{\centering\arraybackslash} m{1.8cm}
			>{\centering\arraybackslash} m{1.8cm}
			>{\centering\arraybackslash} m{1.8cm}
		}
		\toprule
		\multirow{2}{*}{\parbox[c]{1.8cm}{\centering Scenarios}}
		& \multirow{2}{*}{\parbox[c]{1.2cm}{\centering Route name}}
		& \multirow{2}{*}{\parbox[c]{1.5cm}{\centering Tx array height (m)}}
		& \multirow{2}{*}{\parbox[c]{2.4cm}{\centering Horizontal offset angle ($^\circ$)}}
		& \parbox[c]{2.1cm}{\centering Distance between\\ adjacent Rx\\ points (m)}
		& \parbox[c]{1.8cm}{\centering 3D distance between Tx and Rx (m)}
		& \multirow{2}{*}{\parbox[c]{1.8cm}{\centering Number of LoS points}}
		& {\parbox[c]{1.8cm}{\centering Number of foliage shaded points}} \\
		
		\midrule
		
		\multirow{5}{*}{\parbox[c]{1.8cm}{\centering Near-field\\outdoor street}}
		& N1 L1--L7
		& 14.7
		& \parbox[c]{2.2cm}{\centering 45, 30, 15, 0,\\ -15, -30, -45}
		& 3
		& 21--29
		& 21
		& 7 \\[3 mm]
		
		& N1 S1--S7
		& 14.7
		& \parbox[c]{2.2cm}{\centering -6.7, 6.4, 21.8, 10,\\ -21.8, -30.3, 23.4}
		& -
		& 24--29
		& 7
		& 0 \\[3 mm]
		
		& Route1
		& 14.7
		& 0
		& 6
		& 29--42
		& 11
		& 0 \\[2 mm]

		\parbox[c]{1.8cm}{\centering Far-field\\foliage shaded}
		& Route2
		& 16.5
		& 0
		& 3
		& 44--70
		& 0
		& 10 \\[4 mm]

		\parbox[c]{1.8cm}{\centering Far-field LoS\\street canyon}
		& Route3
		& 16.5
		& 0
		& 6
		& 73--222
		& 26
		& 0 \\		
		\bottomrule
		
	\end{tabular}
	\vspace{-1 em}
\end{table*}
\subsection{Near-field Outdoor Street Scenario and Measurement Deployment}
To comprehensively investigate channel characteristics under near-field propagation conditions, channel measurements are conducted on both the N1 horizontal equidistant arcs and route1. In the near-field outdoor street scenario, vegetation and buildings are sparsely distributed on both sides of the road, and metallic scatterers such as street lamps and litter bins are present. The distribution of measurement points is indicated by the orange markers in Fig. \ref{fig:UMi scenario}, while photographs of the near-field LoS and near-field foliage shaded measurement scenarios are shown in Fig. \ref{fig:The photography}\subref{fig:NF LOS} and \ref{fig:The photography}\subref{fig:NF Foliage}.

First, the yellow triangular marker Tx1 denotes the Tx position, with the Tx array height set to 14.7 m and the Rx array height set to 1.8 m. Specifically, within the N1 measurement scenario, seven contiguous sub-routes, denoted L1--L7, are deployed with a horizontal offset angle spacing of 15$^\circ$, covering the range from 45$^\circ$ to -45$^\circ$. Four measurement points are deployed on each of the sub-routes L1--L7 with a horizontal spacing of 3 m. The horizontal distance between the first and fourth points on each sub-route relative to the Tx ranges from 17 m to 26 m. At the same time, seven near-field points are shaded by vegetation with different densities, namely Rx1--Rx2 on sub-route L1, Rx1 on sub-route L2, Rx1 on sub-route L3, and Rx2--Rx4 on sub-route L7. Subsequently, seven points S1--S7 are randomly deployed between sub-routes L1--L7. Finally, route1 is deployed perpendicular to the Tx array coverage direction, with a point spacing of 6 m, comprising 11 points in total. Because Rx4 on sub-route L4 coincides with Rx6 on route1, the total number of near-field measurement points is 46. For sub-routes L1 to L7 and random points S1--S7, Rx array P1 is perpendicular to the direction of the horizontal offset angle extension. For points on route1, the Rx array P1 is parallel to the route direction.

Compared with the discrete-point or linear measurement deployments reported in \cite{VAA100x2510GHzVNA}, \cite{VAA16x2566GHzPN3}, the near-field point deployment method proposed in this paper aligns with the sectorised coverage characteristics of actual cellular base stations, enabling targeted sampling of spatial channels within the real coverage area. Furthermore, by incorporating foliage shaded points and random points, the method further enhances its ability to characterise complex near-field environments and actual user distributions.

\subsection{Far-field Foliage Shaded Scenario and Measurement Deployment}
To investigate the effect of far-field foliage obstruction on signal attenuation, the foliage-shaded scenario along the extension of near-field sub-route L4 is selected. Measurement point distribution is shown along route2 in Fig. \ref{fig:UMi scenario}, with the measurement scenario depicted in Fig. \ref{fig:The photography}\subref{fig:FF Foliage}. Along route2, adjacent Rx points are spaced at 3 m intervals, with a total of 10 measurement points deployed. The 3D distance range between the Tx and Rx arrays is 44--70 m. The Tx array height is 16.5 m, and the Rx array height is 1.8 m.
\par At each measurement point along route2, the Rx array P1 is oriented towards the Tx array, with multiple layers of foliage obstruction of varying densities present between the Rx and Tx arrays. Specifically, Rx1--Rx6 are obstructed by foliage of varying densities relative to the Tx array. Rx7--Rx9 exhibit denser vegetation coverage on both sides than Rx1. Beyond Rx7, the measurement points progressively enter the LoS region, where the vegetation coverage gradually decreases.

\subsection{Far-field LoS Street Canyon Scenario and Measurement Deployment}
To investigate the channel characteristics in the far-field LoS street canyon scenario, 26 measurement points are deployed along route3.  The distribution of the measured points is shown in route3 in Fig. \ref{fig:UMi scenario}, with the actual measurement scenario depicted in Fig. \ref{fig:The photography}\subref{fig:FF LOS}. Along route3, the Tx array height is 16.5 m, and the Rx array height is 1.8 m. The Rx array is positioned 25 m from the buildings at either end, with the second array planar 2 (P2) and the fourth array planar 4 (P4) situated within a densely vegetated environment approximately 5 m from the surrounding trees. Throughout the measurements, the Rx array P1 consistently faced the Tx array, with array P1 oriented perpendicular to route3. The spacing between Rx points is 6 m, and the 3D distance range between the Rx and Tx arrays is 73--222 m. 
\par Compared with the 16--36.5 m measurement range described in the UMi scenario in \cite{VAA100x2510GHzVNA}, and the 50--200 m measurement range in the UMa scenario in \cite{VAA64x12813GHzPN2}, the measurement range in this paper is 21--222 m, covering near-field LoS, far-field foliage-shaded, and far-field LoS propagation conditions. This helps improve the applicability of the channel model in real-world urban scenarios and provide more reliable support for base station coverage assessment, beam design, and long-distance link performance analysis. A summary of measurement-point deployment in near-field outdoor street scenario, far-field foliage shaded scenario, and far-field LoS street canyon scenario is presented in Table \ref{tab:measurement deployment}.

\subsection{Post-Processing of Channel Measurement Data}
\subsubsection{Channel Impulse Response}
To obtain an accurate channel impulse response (CIR), OTA calibration \cite{OTA1} is employed. Compared with direct-connect VNA channel calibration, OTA calibration avoids physical contact, thereby reducing the risk of channel damage caused by manual operations and random phase variations introduced by cable handling. Moreover, it captures practical effects such as antenna gain, array mutual coupling, and the real electromagnetic environment, leading to calibration results that better reflect actual measurements.

First, in the anechoic chamber, the calibration data ${y_{{\rm{cal}}}}\left( \tau  \right)$ are obtained for a known separation between the transceiver array elements, which can be expressed as
\begin{equation}
{y_{{\rm{cal}}}}\left( \tau  \right) = pn\left( \tau  \right) * {g_{{\rm{sys}}}}\left( \tau  \right) * {g_{{\rm{Tx}}}}\left( \tau  \right) * {h_{{\rm{ane}}}}\left( \tau  \right) * {g_{{\rm{Rx}}}}\left( \tau  \right), \label{x_cal}
\end{equation}
where ``$*$'' is the time domain convolution operator, $pn\left( \tau \right)$ denotes the PN sequence, $g_{{\rm{sys}}}\left( \tau  \right)$ denotes the system response, $g_{{\rm{Tx}}}\left( \tau  \right)$ and $g_{{\rm{Rx}}}\left( \tau  \right)$ indicate the transmitter and receiver responses, respectively. $h_{{\rm{ane}}}\left( \tau  \right)$ denotes the anechoic-chamber air-port channel impulse response, which can be expressed as
\begin{equation}
	{h_{{\rm{ane}}}} = \frac{c}{{4\pi d_{\rm {ane}} {f_c}}}\delta \left( {\tau  - {\tau _0}} \right){e^{ - j2\pi {f_c}{\tau _0}}}, \label{h_anechoic}
\end{equation}
where ${c}$ represents the speed of light, ${d_{\rm {ane}}}$ denotes the antenna distance between the Rx array and Tx array, ${{f_c}}$ denotes the centre frequency, and ${{\tau _0}}$ is the propagation delay. Similarly, the outfield air-port test received signal $y\left( \tau  \right)$ can be expressed as
\begin{equation}
y\left( \tau  \right) = pn\left( \tau  \right) * g_{{\rm{sys}}}\left( \tau  \right) * {g_{{\rm{Tx}}}}\left( \tau  \right) * h\left( \tau  \right) * {g_{{\rm{Rx}}}}\left( \tau  \right), \label{outfield air port test receive signal}
\end{equation}
where $h\left( \tau  \right)$ is the measured channel impulse response. Taking the Fourier transform of \beqref{x_cal} and \beqref{outfield air port test receive signal}, we can obtain the frequency domain channel transfer function, which can be expressed as
\begin{equation}
	{Y_{{\rm{cal}}}}\left( f \right) = PN\left( f \right){G_{{\rm{sys}}}}\left( f \right){G_{{\rm{Tx}}}}\left( f \right){H_{\rm {ane}}}\left( f \right){G_{{\rm{Rx}}}}\left( f \right), \label{XX}
\end{equation}
and
\begin{equation}
	Y\left( f \right) = PN\left( f \right){G_{{\rm{sys}}}}\left( f \right){G_{{\rm{Tx}}}}\left( f \right)H\left( f \right){G_{{\rm{Rx}}}}\left( f \right). \label{XX}
\end{equation}

Based on the correlation of the PN sequence and taking the inverse fast Fourier transform (IFFT) of $H\left( f \right)$, we can obtain the channel impulse response $h\left( \tau  \right)$ as
\begin{equation}
	\begin{array}{l}
		{h}\left( \tau  \right) = IFFT\left( {{H}\left( f \right)} \right)\\
		\;	\qquad = IFFT\left( {{{H}_{\rm {ane}}}\left( f \right)\frac{{{Y}\left( f \right){{Y}_{{\rm{cal}}}}^ * \left( f \right)}}{{\left\| {{{Y}_{{\rm{cal}}}}\left( f \right)} \right\|_2^2}}} \right),
	\end{array} \label{(ht)}
\end{equation}
where ${H_{\rm ane}}\left( f \right)$ is the anechoic chamber air-port channel frequency response. ${\left(  \cdot  \right)^ * }$ is a conjugate operation, $\left\|  \cdot  \right\|_2^2$ denotes the operation of squaring the $\ell_2$ norm. Consider a typical MIMO system in which a transmitter with $N_{\rm Tx}$ antennas sends $N_{\rm ds}$ data streams to a receiver with $N_{\rm Rx}$ antennas, and the received signal \textbf {y} can be expressed as
\begin{equation}
	\mathbf{y}=\mathbf{Hs}+\mathbf{n}, \label{XX}
\end{equation}
where \textbf{H} is the channel matrix, \textbf {s} is the signal vector, and \textbf {n} denotes additive noise. We focus on the single-user downlink channel matrix \textbf{H}, and define $\mathbf{H}\left( j,k \right)\in {{\mathbb{C}}^{{{N}_{\text{Rx}}}\times {{N}_{\text{Tx}}}}}$ as the sampled form of $\mathbf{H}\left( t,f \right)\in {{\mathbb{C}}^{{{N}_{\text{Rx}}}\times {{N}_{\text{Tx}}}}}$, where \emph{j} denotes the snapshot ($j=1,2,\ldots ,{{N}_{s}}$) and \emph{k} denotes frequency index ($k=1,2,\ldots {{N}_{f}}$). ${{N}_{s}}$ and ${{N}_{f}}$ represent the total number of snapshots and frequency points, respectively. The time-domain channel impulse response and channel transfer function between the $q$-th receiving antenna and the $p$-th transmitting antenna can be expressed respectively as ${{\emph{h}}_{q,p}}\left( t \right)$ and ${{\emph{H}}_{q,p}}\left(f \right)$.
\begin{figure}[t]
	\centering
	\begin{minipage}[t]{0.49\columnwidth}
		\centering
		\subfloat[]{%
			\includegraphics[width=\linewidth]{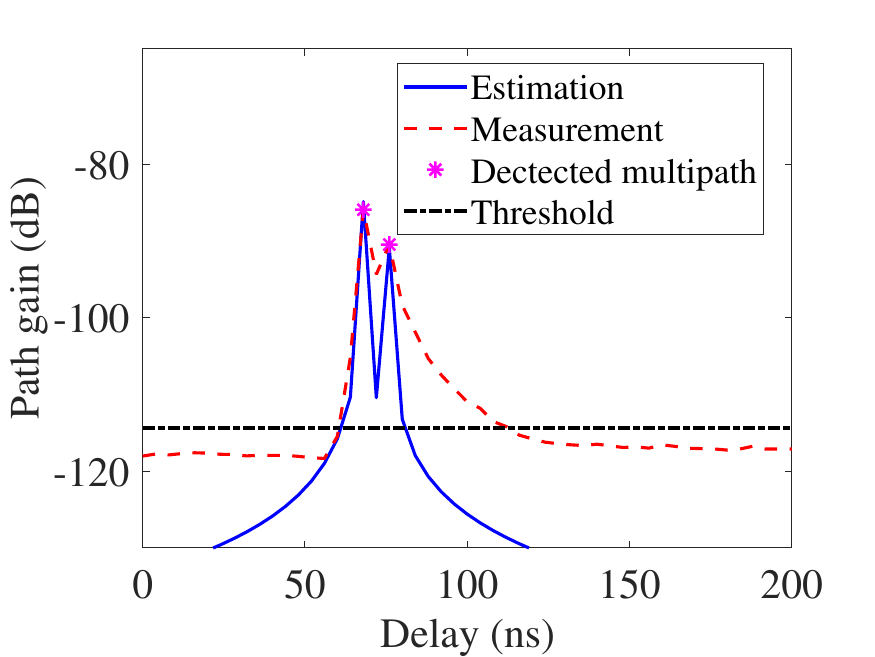}%
			\label{fig:subfig1}%
		}
	\end{minipage}
	\hfill
	\begin{minipage}[t]{0.49\columnwidth}
		\centering
		\subfloat[]{%
			\includegraphics[width=\linewidth]{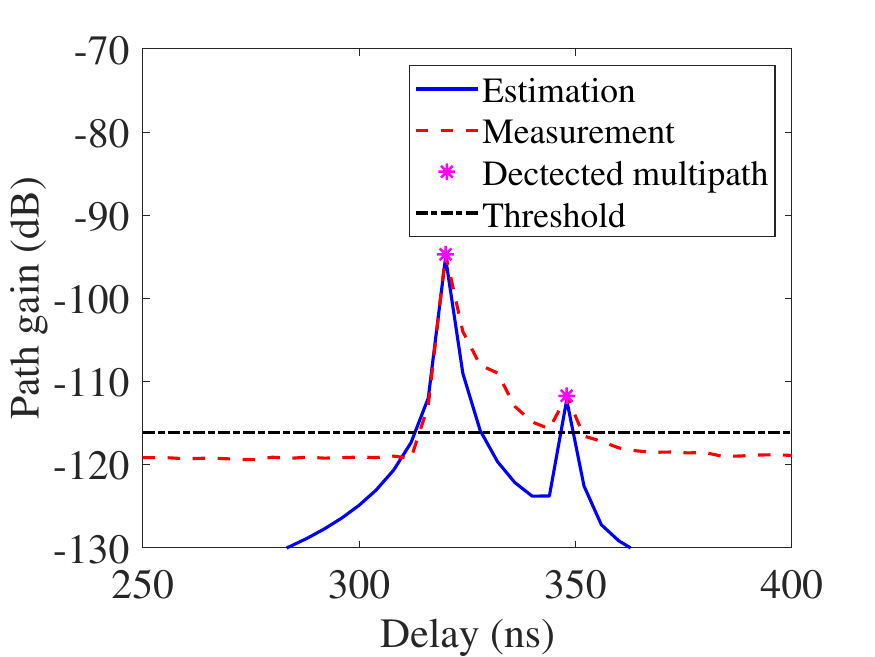}%
			\label{fig:subfig2}%
		}
	\end{minipage}
	
	\caption{Measured and estimated CIRs for array P1 at (a) near-field N1 L6 Rx1 and (b) far-field LoS Rx5.} 
	\label{fig:PDP at N1_L4_Rx1 and LOS Rx5}
\end{figure}
\subsubsection{Channel Multipath Parameters Estimation} 
Owing to its high accuracy and high-resolution performance, the space-alternating generalized expectation-maximization (SAGE) algorithm \cite{SAGE1} has been widely used for multipath component (MPC) parameter estimation in MIMO and wireless propagation channels \cite{NMB3}, \cite{VAA40x4013-17GHzVNA}, \cite{VAAMulti-frequencyVNA}, \cite{VAA16x2566GHzPN3}, \cite{VAA64x12813GHzPN1}, \cite{VAA64x12813GHzPN2}. The MPCs include the complex amplitude ${\alpha _l}$, time delay ${\tau _l}$, azimuth angle of departure (AAoD) ${\phi _{{\rm{Tx,}}l}}$, azimuth angle of arrival (AAoA) ${\phi _{{\rm{Rx}},l}}$, elevation angle of departure (EAOD) ${\theta _{{\rm{Tx}},l}}$, and elevation angle of arrival (EAOA) ${\theta _{{\rm{Rx}},l}}$. To facilitate the extraction of near-field channel MPCs, we employ a subarray partitioning approach \cite{Subarray1}, \cite{Subarray2} for parameter extraction. In the near-field, the SAGE antenna window is set to 16 $\times$ 32. In the far-field, the SAGE antenna window is set to 16 $\times$ 128.

\par Fig. \ref{fig:PDP at N1_L4_Rx1 and LOS Rx5} shows the CIR results obtained by actual measurement and estimation using the SAGE algorithm at near-field N1 L6 Rx1 and the far-field LoS Rx5. Importantly, it can be observed that in the high-power region, the estimated delay and power are highly consistent with the measured results, and the multipath components can be accurately extracted.

\section{Traditional Channel Characterisation Analysis}\label{sec:Traditional}
\subsection{Path Loss Modeling}
\par To investigate the large-scale fading characteristics of UM-MIMO channels at different distances and in various scenarios, path loss modeling and analysis are conducted. Specifically, the near-field LoS scenario comprises the LoS points in N1, points S1--S7, and route1 points. The near-field foliage shaded scenario consists of seven foliage shaded points in N1. The far-field foliage shaded scenario comprises route2 points. The far-field LoS scenario comprises route3 points. 
Furthermore, in the LoS scenarios, the traditional close-in (CI) model and floating intercept (FI) model are considered for path loss modeling. And the COST 235 model is used for the path loss modeling in the foliage shaded scenario.

\subsubsection{CI path loss modeling}
The CI path loss model \cite{NMB3}, \cite{CIXDS}, \cite{CI3} can be expressed as  
\begin{equation}
	P{L^{{\rm{CI}}}}[{\rm{dB}}] = 10 \cdot PLE\cdot{\log _{10}}\left( {\frac{d}{{{d_0}}}} \right) + FSPL\left( {{d_0}} \right) + X_{{\sigma _{{\rm{CI}}}}}^{{\rm{CI}}}, \label{(CI path loss modeling)}
\end{equation}
where $PLE$ is the path loss exponent, ${d}$ denotes the antenna array distance between the Rx array and Tx array, ${{d_0}}$ is the reference distance which is set to 1 m. $X_{{\sigma _{{\rm{CI}}}}}^{{\rm{CI}}}$ is a normally distributed random variable with the standard deviation ${{\sigma _{{\rm{CI}}}}}$. $FSPL\left( {{d_0}} \right)$ stands for the free-space path loss at a reference distance ${{d_0}}$ and can be expressed as
${FSPL\left( {{d_0}} \right) =  - 20\cdot{\log _{10}}\left( {\frac{c}{{4\pi f{d_0}}}} \right)}$, and ${f}$ stands for the carrier frequency.
\subsubsection{FI path loss modeling}
In \beqref{(CI path loss modeling)}, it can be seen that the intercept of the CI path loss model is the free-space path loss at the reference distance, which may limit the applicability of the model. In order to fit the path loss scenario better at different frequencies and under different scenarios, the FI path loss model \cite{CIXFI} is considered, which can be expressed as
\begin{equation}
	P{L^{{\rm{FI}}}}\left[ {{\rm{dB}}} \right] = {\alpha _{{\rm{FI}}}} + 10\cdot{\beta _{{\rm{FI}}}}\cdot{\log _{10}}\left( {\frac{d}{{{d_0}}}} \right) + X_{{\sigma _{{\rm{FI}}}}}^{{\rm{FI}}}, \label{XX}
\end{equation}
where ${\alpha _{{\rm{FI}}}}$ and ${\beta _{{\rm{FI}}}}$ represent the floating intercept and the FI path loss exponent, respectively, and $X_{{\sigma _{{\rm{FI}}}}}^{{\rm{FI}}}$ denotes a normally distributed random variable with the standard deviation of ${{\sigma _{{\rm{FI}}}}}$.

\subsubsection{COST 235 path loss modeling}
To account for the effect of different vegetation depths ${{d_{\rm{v}}}}$ on path loss, the path loss in the case of foliage shading can be expressed as
\begin{equation}
	P{L^{{\rm{fol}}}}\left( {d,{d_{\rm{v}}}} \right) = FSPL\left( d \right) + {L_{\rm{v}}}\left( {{d_{\rm{v}}}} \right), \label{XX}
\end{equation}
where ${L_{\rm{v}}}\left( {{d_{\rm{v}}}} \right)$ denotes the additional path loss due to branch and leaf attenuation. The additional path loss model adopted in COST 235 \cite{COST235}, which can be expressed as
\begin{equation}
	{L_{{\rm{v - COST235}}}}\left( {{d_{\rm{v}}}} \right) = Af_{\rm{v}}^Bd_{\rm{v}}^C, \label{XX}
\end{equation}
where $A$, $B$, and $C$ represent the model parameters with respect to amplitude, frequency, and vegetation depth, respectively, the units of ${f_{\rm{v}}}$ and ${d_{\rm{v}}}$ are MHz and metres, respectively. In the COST 235 model, the path loss under foliage shaded can be categorised into two scenarios, out of leaf and in leaf, which can be represented as ${L_{{\rm{v - COST235 - out}}}} = 26.6f_{\rm{v}}^{ - 0.2}d_{\rm{v}}^{0.5}$ and ${L_{{\rm{v - COST235 - in}}}} = 15.6f_{\rm{v}}^{ - 0.009}d_{\rm{v}}^{0.26}$, respectively.
\subsection{Omni-Directional Path Loss}
As a large-scale fading parameter, path loss plays an important role in the link budget and system design of communication systems. Because OTA calibration is employed, the received signal $y\left( \tau  \right)$ measured in the UMi environment is calibrated against a microwave anechoic chamber reference signal ${y_{{\rm{cal}}}\left( \tau  \right)}$ to obtain the true path loss. During calibration in a microwave anechoic chamber, the measured path loss can be expressed as
\begin{equation}
	\begin{split}
		P{L^{{\rm{ane}}}}\left[ {{\rm{dB}}} \right] ={}& {P_{{\rm{VSG}}}} + G_{{\rm{Tx}}}^{{\rm{PA}}} + G_{{\rm{Tx}}}^{{\rm{Ant}}} + G_{{\rm{Rx}}}^{{\rm{LNA}}} + \\
		& G_{{\rm{Rx}}}^{{\rm{Ant}}} - P_{\rm{r}}^{{\rm{ane}}} - {P_{{\rm{CL}}}}, \label{(PL_A)}
	\end{split}
\end{equation}
where the cable loss ${P_{{\rm{CL}}}}$ consists the loss from a 7-metre RF cable connecting the VSG to the Tx antenna array, and the loss from a 2-metre RF cable connecting the Rx antenna array to the SA. $P_{\rm{r}}^{{\rm{ane}}}$ denotes the signal power received by the SA in the microwave anechoic chamber.
\begin{figure}[t]
	\centering
	\subfloat[]{%
		\includegraphics[width=0.50\columnwidth]{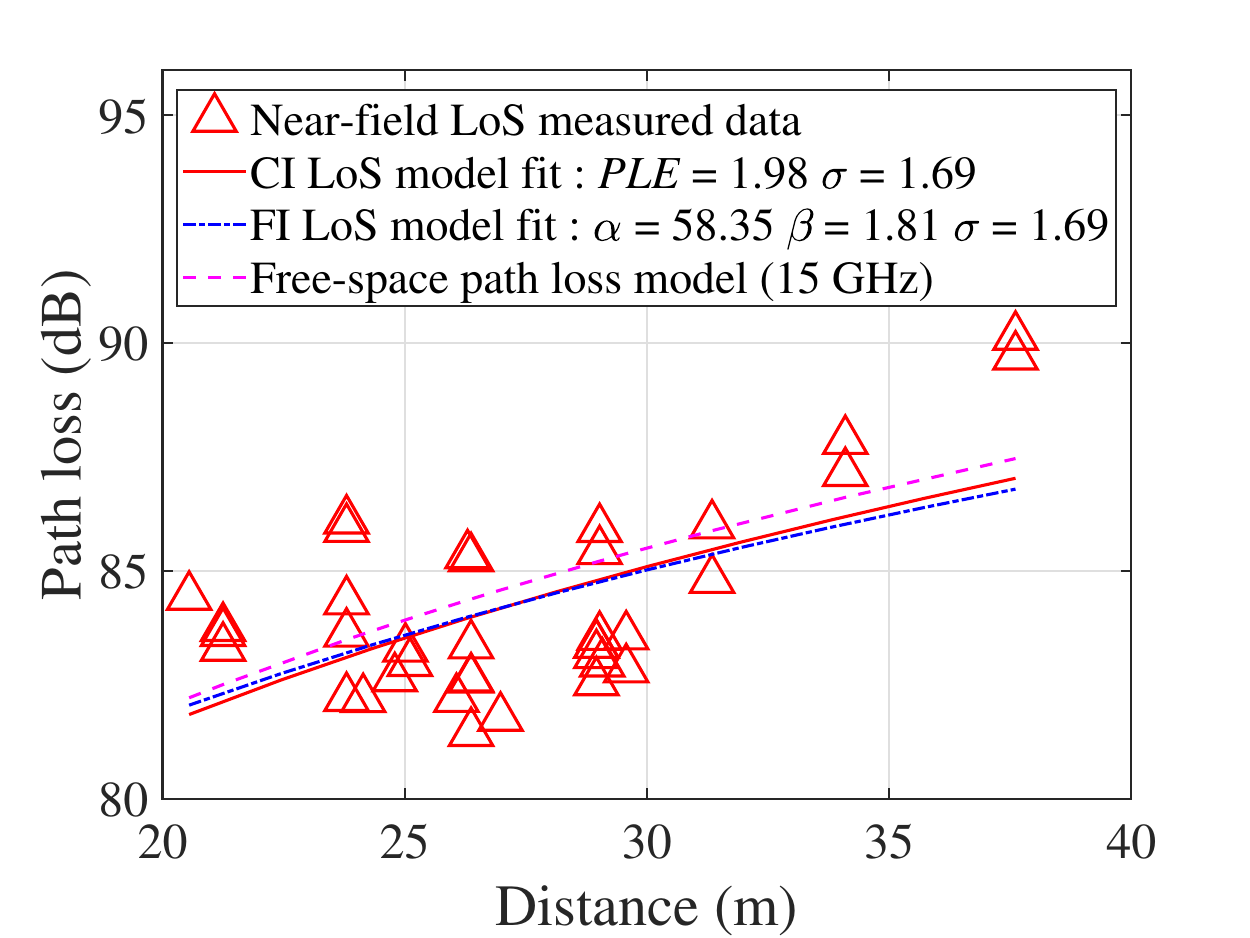}%
		\label{fig:subfig1}%
	}
	\subfloat[]{%
		\includegraphics[width=0.50\columnwidth]{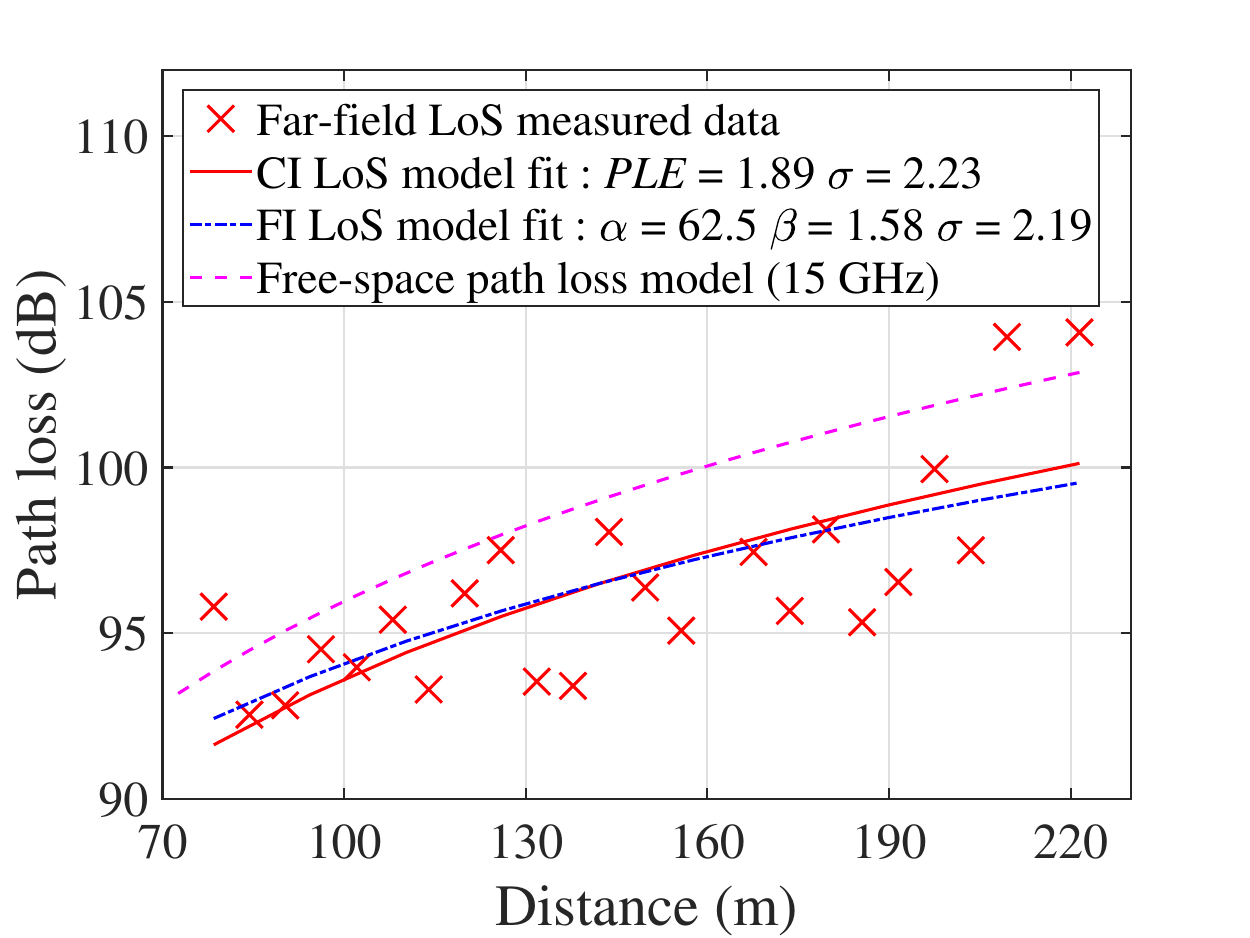}%
		\label{fig:subfig2}%
	}
	\caption{Omni-directional path loss measurement results and single-frequency CI model and FI model in (a) near-field LoS and (b) far-field LoS scenarios at 15 GHz.} 
	\label{fig:PL_NF_FF_LOS}
\end{figure}
\par During channel measurements in the UMi environment, the path loss obtained using identical equipment and cables can be expressed as \begin{equation}
	\begin{split}
		P{L^{{\rm{mea}}}}\left[ {{\rm{dB}}} \right] ={}& {P_{{\rm{VSG}}}} + G_{{\rm{Tx}}}^{{\rm{PA}}} + G_{{\rm{Tx}}}^{{\rm{Ant}}} +  G_{{\rm{Rx}}}^{{\rm{LNA}}} + \\ & G_{{\rm{Rx}}}^{{\rm{Ant}}} - P_{\rm{r}}^{{\rm{mea}}} - {P_{{\rm{CL}}}}, \label{(PL_M)}
	\end{split}
\end{equation}
where $P_{\rm{r}}^{{\rm{mea}}}$ denotes the received signal power of the SA in the actual UMi measurement environment.
\par Because an anechoic chamber can be regarded as a free-space propagation environment, $P{L^{{\rm{ane}}}} = FSPL\left( {{d_{{\rm{ane}}}}} \right)$. According to \beqref{(PL_A)} and \beqref{(PL_M)}, the measured path loss can be obtained as
\begin{equation}
	\begin{split}
		P{L^{{\rm{mea}}}}\left[ {{\rm{dB}}} \right] ={} - \left( {P_{\rm{r}}^{{\rm{mea}}} - P_{\rm{r}}^{{\rm{ane}}}} \right) +   FSPL\left( {{d_{{\rm{ane}}}}} \right). \label{(PL_omni)}
	\end{split}
\end{equation}
\par According to \beqref{(ht)}, the raw calibrated channel is obtained as
\begin{equation}
	\begin{split}
		{H_{{\rm{raw}}}}\left( f \right) = \frac{{Y\left( f \right)}}{{{Y_{{\rm{cal}}}}\left( f \right)}} = \frac{{H\left( f \right)}}{{{H_{\rm ane}}\left( f \right)}},
	\end{split}
\end{equation}
taking the inverse Fourier transform of ${H_{{\rm{raw}}}}\left( f \right)$ yields ${h_{{\rm{raw}}}}\left( \tau  \right) = \sum\limits_l {{\alpha _l}} \delta \left( {\tau  - {\tau _l}} \right)$ , ${\tau _l}$ and ${\alpha _l}$ denote the delay and the complex gain of the $l$-th multipath, respectively.
\par For the raw calibrated channel ${H_{{\rm{raw}}}}\left( f \right)$, the SAGE algorithm is employed to estimate distinguishable multipath complex gains ${\alpha _l}$ with varying delays, angles of arrival (AOA), and angles of elevation (EOA). During the multipath extraction process in the SAGE algorithm, the selected noise threshold is $\max \left( {{P_{\rm{m}}} - 30,{P_{\rm{n}}} + 3} \right)$, ${P_{\rm{m}}}$ is the peak power, and ${P_n}$ denotes noise power. Ultimately, the omni-directional received power is obtained by summing the powers of all unique paths in the power domain, and can be expressed as
\begin{equation}
	\begin{split}
		P_{{\rm{Rx}}}^{{\rm{Omni}}} {}= P_{\rm{r}}^{{\rm{mea}}} - P_{\rm{r}}^{{\rm{ane}}} =  10 \cdot {\log _{10}}\left( {\sum\limits_{l = 1}^L {{{\left| {{\alpha _l}} \right|}^2}} } \right), \label{(P_Rx_omni)}
	\end{split}
\end{equation}
where $L$ denotes the total number of distinguishable multipaths estimated by the SAGE algorithm. By substituting \beqref{(P_Rx_omni)} into  \beqref{(PL_omni)}, the measured omni-directional path loss is obtained.

\subsection{Near-Field and Far-Field LoS Path Loss}
Fig. \ref{fig:PL_NF_FF_LOS} shows the omni-directional path loss measurements and the fitting results obtained using the CI model and FI model for near-field LoS and far-field LoS scenarios, respectively. The fitting results are shown in Table \ref{tab:PL_table}. In the CI model, the PLE values of near-field LoS and far-field LoS are 1.98 and 1.89, respectively, both of which are smaller than the free-space loss value of 2, indicating that the measured path loss value meets the LoS propagation condition. 
For the near-field LoS channel path loss modeling, the root mean square error (RMSE) values between the measured path loss and the CI and FI models are 1.671 dB and 1.668 dB, respectively. This indicates that both models fit the measured data well, although the FI model shows slightly better accuracy. Similarly, in the far-field LoS channel path loss modeling, the RMSE values of the CI and FI models are 2.18 dB and 2.14 dB, respectively, which again suggests that the FI model achieves higher accuracy than the CI model.
\begin{figure}[t]
	\centering
	\subfloat[]{%
		\includegraphics[width=0.50\columnwidth]{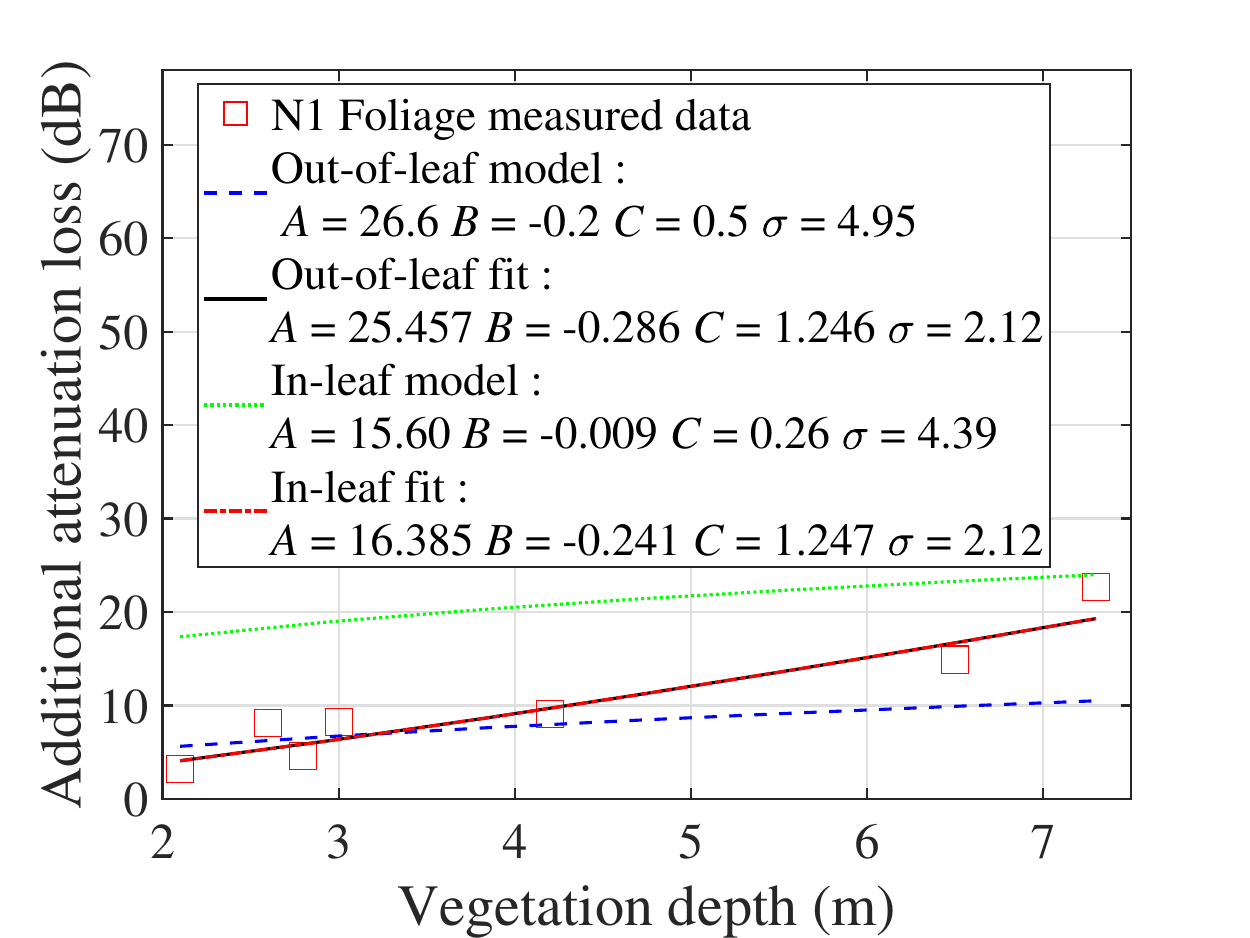}%
		\label{fig:NF foliage shaded}%
	}
	\subfloat[]{%
		\includegraphics[width=0.50\columnwidth]{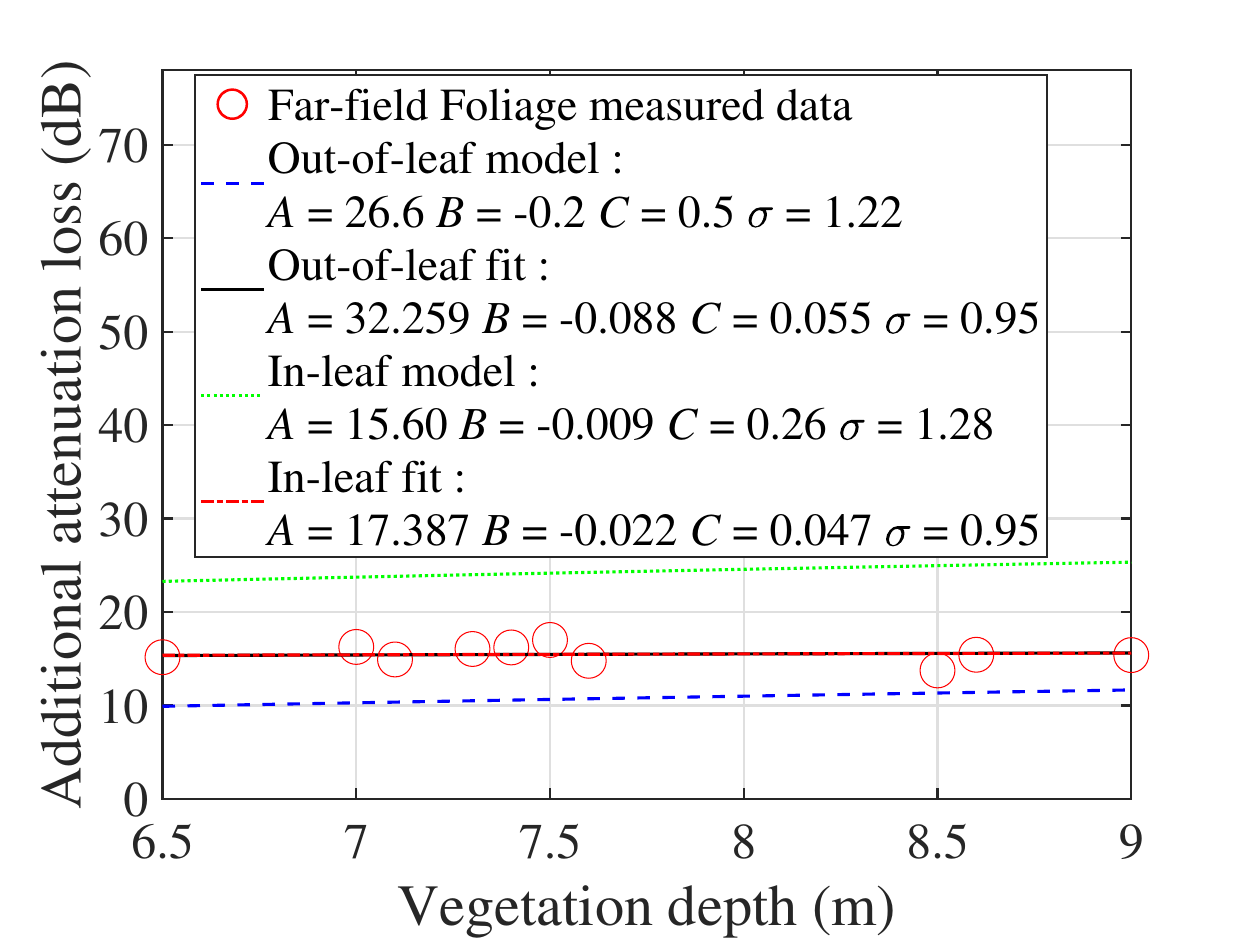}%
		\label{fig:FF foliage shaded}%
	}
	\caption{Additional attenuation loss measurement results and COST 235 models in (a) near-field foliage shaded and (b) far-field foliage shaded scenarios at 15 GHz} 
	\label{fig:PL_NF_FF_foliage}
\end{figure}
\subsection{Near-field and Far-field Foliage Shaded Additional Path Loss}

Fig. \ref{fig:PL_NF_FF_foliage} shows the additional loss measurements and the results of model fitting using COST
235 for the near-field foliage and far-field foliage scenarios, respectively. The model fitting results are shown in Table \ref{tab:PL_table}. It is clear that in the near-field region, where vegetation depth is less than 3 metres, the average additional loss is about 7 dB. In the far-field foliage shaded scenario, however, the average additional loss is about 16 dB. This is likely because the points in the near-field foliage-shaded scenario are closer to the Tx, resulting in higher received power. Specifically, in the near-field foliage shaded scenario, the RMSE values of the original COST 235 out-of-leaf and in-leaf models relative to the measured values are 5.21 dB and 10.87 dB, respectively. The RMSE values of the COST 235 out-of-leaf and in-leaf fitting models relative to the measured values are both 2.03 dB. Similarly, in the far-field foliage shaded scenario, the RMSE values of the original COST 235 out-of-leaf and in-leaf models relative to the measured values are 4.89 dB and 8.85 dB, respectively. The RMSE values of the COST 235 out-of-leaf and in-leaf fitting models relative to the measured values are 0.91 dB and 0.90 dB, respectively. This indicates that the COST 235 out-of-leaf and in-leaf fitting models are more accurate.
\subsection{Delay and Angle Domain Channel Characteristics}
To investigate the characteristics of UM-MIMO channels in the delay and angular domains under different scenarios, the power delay angle profile (PDAP), the root mean square delay spread (RMS DS) and the angular spread (AS) are analyzed.
\subsubsection{Power Delay Angle Profiles}
PDAPs can be used to describe the distributions of MPCs in terms of delay and arrival angle across different measurement scenarios, thereby providing an intuitive view of channel propagation mechanisms, environmental scattering structures and scenario differences.
\begin{table}[t]
	\renewcommand{\arraystretch}{1} 
	\centering
	\caption{Path Loss Model Coefficients For NF LoS, FF LoS, NF Foliage Shaded and FF Foliage Shaded Scenarios}
	\label{tab:PL_table}
	\begin{tabular}{|>{\centering\arraybackslash}m{1cm}|
			>{\centering\arraybackslash}m{1.6cm}|					 
			>{\centering\arraybackslash}m{0.8cm}|
			>{\centering\arraybackslash}m{0.8cm}|
			>{\centering\arraybackslash}m{0.8cm}|
			>{\centering\arraybackslash}m{0.7cm}|
		}
		\hline
		\multirow{2}{*}{Scenarios} & \multirow{2}{*}{PL Model} &\multicolumn{4}{c|}{Parameters} \\
		\cline{3-6}	
		&  & {\parbox{0.8cm}{\centering $PLE$}} & {\parbox{0.8cm}{\centering $\alpha$}} & {\parbox{0.8cm}{\centering $\beta$ }} & {\parbox{0.7cm}{\centering $\sigma$ }} \\ \hline
		\multirow{2}{*}{\parbox{1.1cm}{\centering Near-field LoS}}
		& {\parbox{1.6cm}{\centering CI}}   &{\parbox{0.8cm}{\centering 1.98}} & -  	 & -   	&{\parbox{0.7cm}{\centering 1.69}}   \\ \cline{2-6}
		& {\parbox{1.6cm}{\centering FI}}   & -    &{\parbox{0.8cm}{\centering 58.35}} &{\parbox{0.8cm}{\centering 1.81}} &{\parbox{0.7cm}{\centering 1.69}}   \\ \cline{2-6}
		\hline
		\multirow{2}{*}{\parbox{1.1cm}{\centering Far-field LoS}}
		& {\parbox{1.6cm}{\centering CI}}   &{\parbox{0.8cm}{\centering 1.89}} & -  & -  	&{\parbox{0.7cm}{\centering 2.23}} \\ \cline{2-6}
		& {\parbox{1.6cm}{\centering FI}}   &-     &{\parbox{0.8cm}{\centering 62.5}} &{\parbox{0.8cm}{\centering 1.58}}      &{\parbox{0.7cm}{\centering 2.19}}  \\ \cline{2-6}
		\hline
		\multicolumn{2}{|c|}{} &{\parbox{0.8cm}{\centering ${A}$}} & {\parbox{0.8cm}{\centering ${B}$}} & {\parbox{0.8cm}{\centering ${C}$}} &{\parbox{0.7cm}{\centering{$\sigma$ }}} \\
		\hline
		\multirow{8}{*}{\parbox{1cm}{\centering Near-field foliage shaded }}
		& COST 235 out-of-leaf  	&{\parbox{0.8cm}{\centering 26.6}}   &{\parbox{0.8cm}{\centering -0.2}}   &{\parbox{0.8cm}{\centering 0.5}}   &{\parbox{0.7cm}{\centering {4.95}}}   \\ \cline{2-6}
		& COST 235 out-of-leaf fit  &{\parbox{0.8cm}{\centering 25.457}} &{\parbox{0.8cm}{\centering -0.286}} &{\parbox{0.8cm}{\centering 1.246}} &{\parbox{0.7cm}{\centering {2.12}}}   \\ \cline{2-6}
		& COST 235 in-leaf  		&{\parbox{0.8cm}{\centering 15.60}}  &{\parbox{0.8cm}{\centering -0.009}} &{\parbox{0.8cm}{\centering 0.26}}  &{\parbox{0.7cm}{\centering {4.39}}}   \\ \cline{2-6}
		& COST 235 in-leaf fit  	&{\parbox{0.8cm}{\centering 16.385}} &{\parbox{0.8cm}{\centering -0.241}} &{\parbox{0.8cm}{\centering 1.247}} &{\parbox{0.7cm}{\centering {2.12}}}   \\ \cline{2-6}
		\hline
		\multirow{8}{*}{\parbox{1cm}{\centering Far-field foliage shaded}}
		& COST 235 out-of-leaf  	&{\parbox{0.8cm}{\centering 26.6}}   &{\parbox{0.8cm}{\centering -0.2}}    &{\parbox{0.8cm}{\centering 0.5}}   &{\parbox{0.7cm}{\centering {1.22}}}  \\ \cline{2-6}
		& COST 235 out-of-leaf fit  &{\parbox{0.8cm}{\centering 32.259}} &{\parbox{0.8cm}{\centering -0.088}}  &{\parbox{0.8cm}{\centering 0.055}} &{\parbox{0.7cm}{\centering {0.95}}}  \\ \cline{2-6}
		& COST 235 in-leaf  		&{\parbox{0.8cm}{\centering 15.60}}  &{\parbox{0.8cm}{\centering -0.009}}  &{\parbox{0.8cm}{\centering 0.26}}  &{\parbox{0.7cm}{\centering {1.28}}}  \\ \cline{2-6}
		& COST 235 in-leaf fit  	&{\parbox{0.8cm}{\centering 17.387}} &{\parbox{0.8cm}{\centering -0.022}}  &{\parbox{0.8cm}{\centering 0.047}} &{\parbox{0.7cm}{\centering {0.95}}}  \\ \cline{2-6}
		\hline
	\end{tabular}
\end{table} 

\par
\begin{figure}[t]
	\centering
	\subfloat[Route1, Rx7.]{%
		\includegraphics[width=0.49\columnwidth]{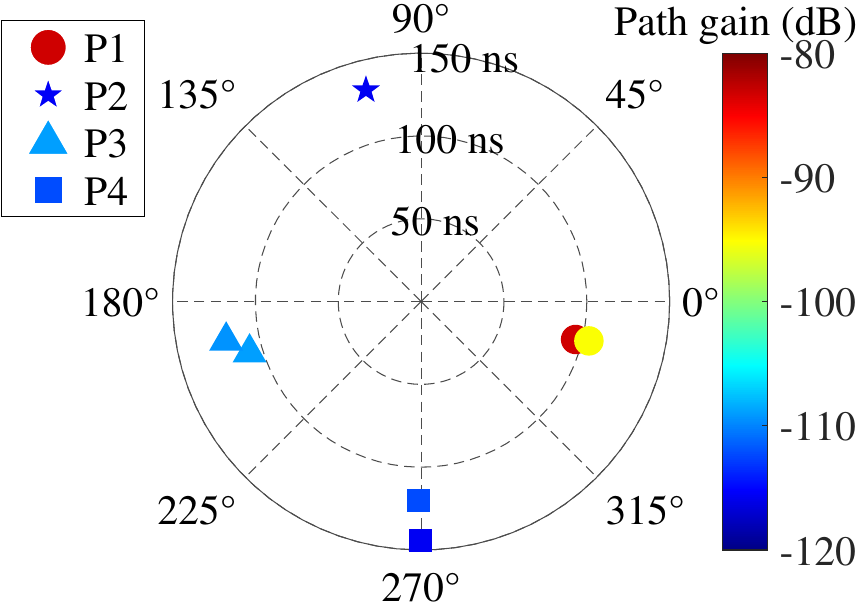}%
		\label{fig:PDAP at route1 Rx3}%
	}
	\hfill
	\subfloat[Route3, Rx5.]{%
		\includegraphics[width=0.49\columnwidth]{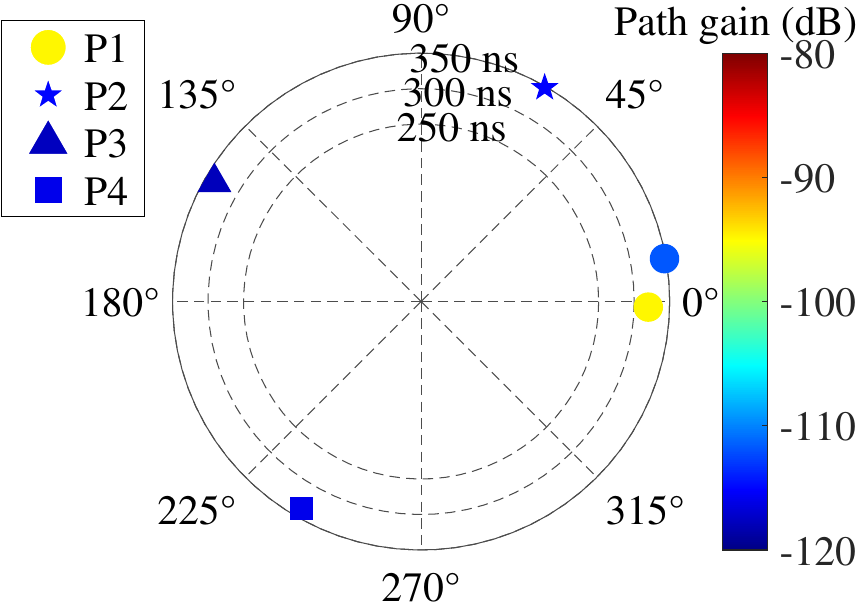}%
		\label{fig:PDAP at route2 Rx11}
	}
	
	\subfloat[N1, L7 Rx4.]{%
		\includegraphics[width=0.49\columnwidth]{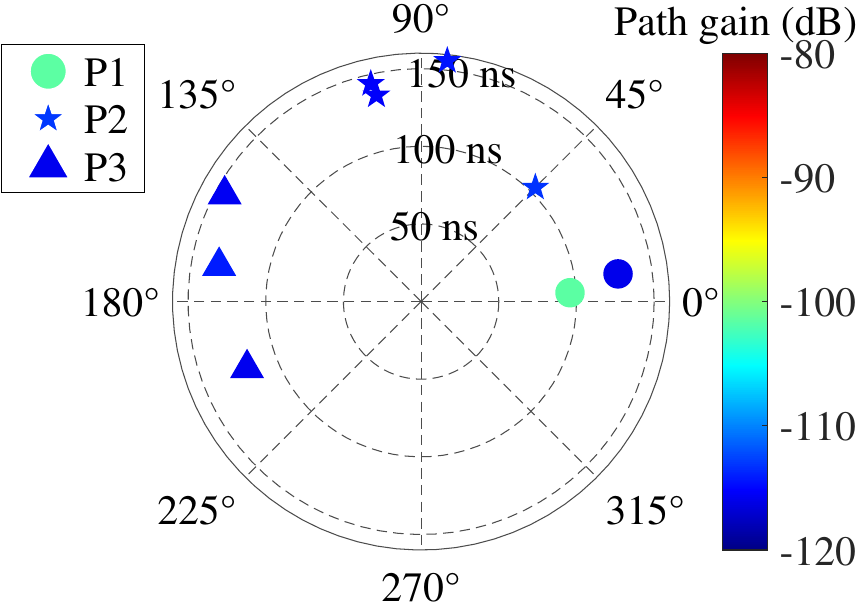}%
		\label{fig:PDAP at route3 Rx1}%
	}
	\hfill
	\subfloat[Route2, Rx7.]{%
		\includegraphics[width=0.49\columnwidth]{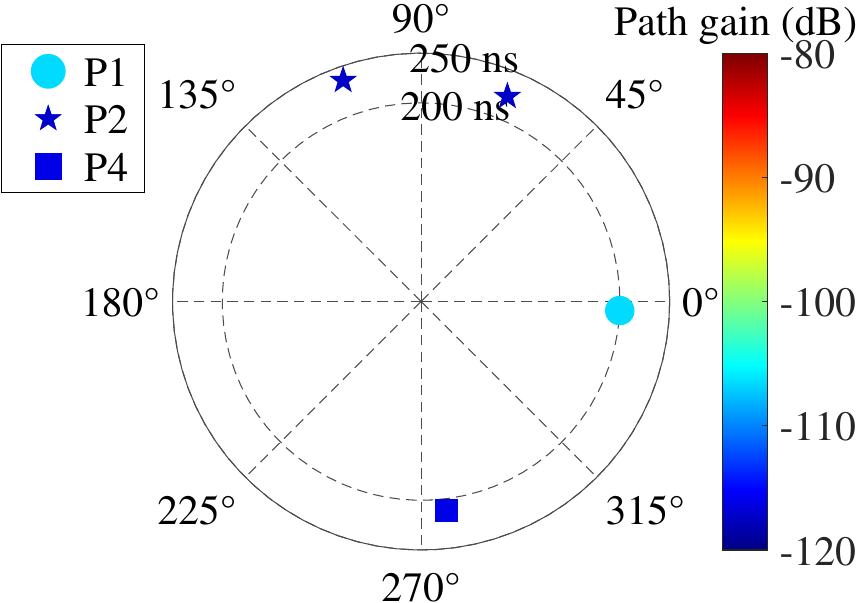}%
		\label{fig:PDAP at route4 Rx1}	
	}
	\caption{The PDAPs at (a) route1, Rx7, (b) route3 Rx5, (c) N1, L7 Rx4, and (d) route2 Rx7.}
	\label{fig:The PDAPs}
\end{figure}

 Fig. \ref{fig:The PDAPs} shows the PDAP results for the LoS and foliage shaded scenarios under far-field and near-field conditions. At the Rx7 on route1, the Rx primarily receives the LoS MPC at P1, with LoS delay around 100 ns and angle of arrival (AoA) concentrated around 346°. Due to reflections from scatterers such as street lamps and litter bins, the Rx receives a small amount of MPCs across the array P2 to array P4. Similarly, at the Rx5 on route3, the Rx mainly receives the LoS MPC at P1, with the LoS MPC delay being approximately 300 ns and the AoA centred around 0°. Due to reflections from scatterers such as street lamps and the building B3, the Rx receives small amounts of MPCs across the array P2 to array P4. At Rx4 on N1 sub-route L7, as the Rx array’s P1 planar is perpendicular to the horizontal propagation direction of L7, the Rx does not receive multipath at the array P4. At Rx7 route2, due to the foliage shaded, there is little multipath at P2 and P4, and no multipath is received at array P3.
\subsubsection{Delay Spread}
As multipath components reach the Rx via different propagation paths, the signal exhibits varying degrees of dispersion in the time domain. This phenomenon is typically described by power delay profiles (PDPs). Furthermore, this time dispersion characteristic is commonly quantified using the RMS DS \cite{CIXDS}, \cite{RMSDS}, which can be expressed as  
\begin{equation}
	{\emph {RMS\,DS}} = \sqrt {\frac{{\sum\limits_{l = 1}^L {\tau _l^2P\left( {{\tau _l}} \right)} }}{{\sum\limits_{l = 1}^L {P\left( {{\tau _l}} \right)} }} - {{\left( {\frac{{\sum\limits_{l = 1}^L {{\tau _l}P\left( {{\tau _l}} \right)} }}{{\sum\limits_{l = 1}^L {P\left( {{\tau _l}} \right)} }}} \right)}^2}},   \label{(RMSDS)}
\end{equation}
where ${P\left( {{\tau _l}} \right)}$ is the power of the $l$-th multipath. 

\par Detailed statistical values of RMS DS for different scenarios, together with the corresponding values recommended in 3GPP TR 38.901 \cite{3GPP38901}, are summarized in Table \ref{table:DS_AS_Table}. The lognormal fit mean value of near-field LoS, far-field LoS, near-field foliage shaded, and far-field (FF) foliage shaded scenarios are -8.50 (3.17 ns), -8.89 (1.28 ns), -8.41 (3.92 ns), and -7.48 (33.43 ns), respectively. Overall, the mean values of RMS DS in LoS scenarios are generally small, with the FF LoS scenario exhibiting the smallest value, indicating that when the LoS main path dominates, multipath energy is distributed more concentratively in the delay domain. 
\par Compared with the NF LoS scenario, the mean value of RMS DS in the NF foliage shaded scenario increased only slightly, indicating that whilst NF foliage obstruction introduces some additional scattering, its impact on delay spread is relatively limited. By contrast, the mean value of RMS DS in the FF foliage shaded scenario is significantly higher than in the other three scenarios, suggesting that FF foliage shaded enhances multipath scattering effects, resulting in a more dispersed signal delay distribution. Furthermore, when compared with the 3GPP UMi LoS reference value, the mean values of RMS DS in both the NF LoS and FF LoS scenarios are lower, indicating that the multipath structure in the measured LoS scenarios is more compact. Simultaneously, the standard deviation in the NF LoS scenario is lower than the reference value, whereas in the FF LoS scenario it is markedly higher than the reference value, indicating that near-field propagation is more stable, whereas far-field propagation is more susceptible to the influence of scatterers such as buildings and vegetation, leading to increased multipath delay jitter.

\begin{table}[t]
	\centering
	\caption{Summaries of Measurement-based RMS DS and AS Versus 3GPP TR 38.901 in the UMi Environment}
	\label{table:DS_AS_Table}
	\setlength{\tabcolsep}{0pt}
	\renewcommand{\arraystretch}{1.0}
	\begin{tabular}{%
			>{\centering\arraybackslash}m{1.6cm}
			>{\centering\arraybackslash}m{1.2cm}
			>{\centering\arraybackslash}m{1cm}
			>{\centering\arraybackslash}m{1cm}
			>{\centering\arraybackslash}m{1cm}
			>{\centering\arraybackslash}m{1cm}
			>{\centering\arraybackslash}m{1cm}
			>{\centering\arraybackslash}m{1cm}
		}
		\toprule
		& & \multicolumn{4}{c}{Measurement} & \multicolumn{2}{c}{3GPP TR 38.901}\\
		\cmidrule(lr){3-6}\cmidrule(lr){7-8}
		\multirow{-2}{*}{\text{Parameters}} & 
		& \shortstack{NF\\LoS}
		& \shortstack{FF\\LoS}
		& \shortstack{NF\\Foliage}
		& \shortstack{FF\\Foliage}
		& LoS
		& Foliage\\
		\midrule
		
		\multirow{2}{*}{\shortstack{DS\\(log$_{10}$ [s])}}
		& $\mu_{\rm{DS}}$    & -8.50 & -8.89 & -8.41 & -7.48 & -7.43 & - \\
		& $\sigma_{\rm{DS}}$ &  0.32 &  1.75 &  0.36 &  0.23 &  0.38 & - \\
		\midrule
		
		\multirow{6}{*}{\shortstack{AS\\(log$_{10}$ [$^\circ$])}}
		& $\mu_{\rm{ASA}}$    & 0.87 & 0.06  & 0.83  & 1.69  & 1.63  & - \\
		& $\sigma_{\rm{ASA}}$ & 0.24 & 2.10  & 0.63  & 0.07  & 0.30  & - \\[1mm]
		& $\mu_{\rm{ASD}}$    & 0.69 & 0.31  & 0.87  & 0.72  & 1.15  & - \\
		& $\sigma_{\rm{ASD}}$ & 0.31 & 2.16  & 0.65  & 2.56  & 0.41  & - \\[1mm]
		& $\mu_{\rm{ESD}}$    & 0.53 &-0.99  & 0.68  & 1.23  & 0.54  & - \\
		& $\sigma_{\rm{ESD}}$ & 0.32 & 2.70  & 0.27  & 0.24  & 0.35  & - \\
		\bottomrule
	\end{tabular}
\end{table}
\subsubsection{Angular Spread}
AS is one of the key parameters characterising small-scale channel properties, and reflects the degree of dispersion of MPCs in the angular domain. In the UM-MIMO system, angular spread typically comprises the arrival azimuth angle spread (ASA), the departure azimuth angle spread (ASD) and the departure elevation angle spread (ESD). AS can be expressed as \cite{3GPP38901} 
\begin{equation}
AS = \sqrt { - 2\ln \left( {\left| {\frac{{\sum\nolimits_{l = 1}^L {exp\left( {j{\varphi _l}} \right){P_l}} }}{{\sum\nolimits_{l = 1}^L {{P_l}} }}} \right|} \right)},   \label{XX}
\end{equation}
where $\varphi_l$ represents either the azimuth angle $\phi_l$ or the elevation angle $\theta_l$ of the $l$-th MPC, and $P_l$ is the power of the $l$-th MPC. A larger AS value indicates a broader angular distribution of MPCs, implying stronger multipath propagation and a richer scattering environment. 

\par The AS statistics and the values recommended in 3GPP TR 38.901 for different scenarios are also summarized in Table \ref{table:DS_AS_Table}. The statistical results indicate that, under LoS conditions, the mean values of ASA, ASD and ESD decrease from 0.87, 0.69 and 0.53 in the near field to 0.06, 0.31 and -0.99 in the far field, respectively. This suggests that as the path transitions from the near field to the far field, the dominant influence of the LoS path further increases, and the angular energy distribution converges markedly, with the most significant contraction occurring in the azimuth dimension. In contrast, vegetation obstruction causes a marked widening of the angular spread. Particularly in the FF foliage shaded scenario, the mean values of ASA and ESD increased to 1.69 and 1.23 respectively, reaching the maximum values among the four scenarios, indicating that scattering, diffraction and transmission effects induced by vegetation significantly enhance the spread in the arrival azimuth and departure elevation directions. At the same time, the standard deviations of all parameters increased markedly in the FF LoS scenario, reflecting greater fluctuations and poorer stability in the angular spread of unobstructed far-field links. 

As can be clearly seen from Table \ref{table:DS_AS_Table}, the current 3GPP TR 38.901 provides only general LoS reference values for the UMi environment. The statistical values for RMS DS and AS under near-field/far-field conditions and vegetation-shaded conditions have not yet been fully reflected. Based on experimental results, this paper further quantifies relevant parameters in near-field/far-field LoS and near-field/far-field foliage-shaded scenarios, and reveals key patterns not reflected in the standard model. These include AS contraction during the transition from near-field to far-field under LoS conditions, as well as a significant increase in DS and AS caused by far-field vegetation obstruction. The results provide experimental evidence for the further refinement of the 3GPP channel model.

\begin{figure}[t]
	\centering
	\subfloat[The LoS path magnitude.]{%
		\includegraphics[width=0.50\columnwidth]{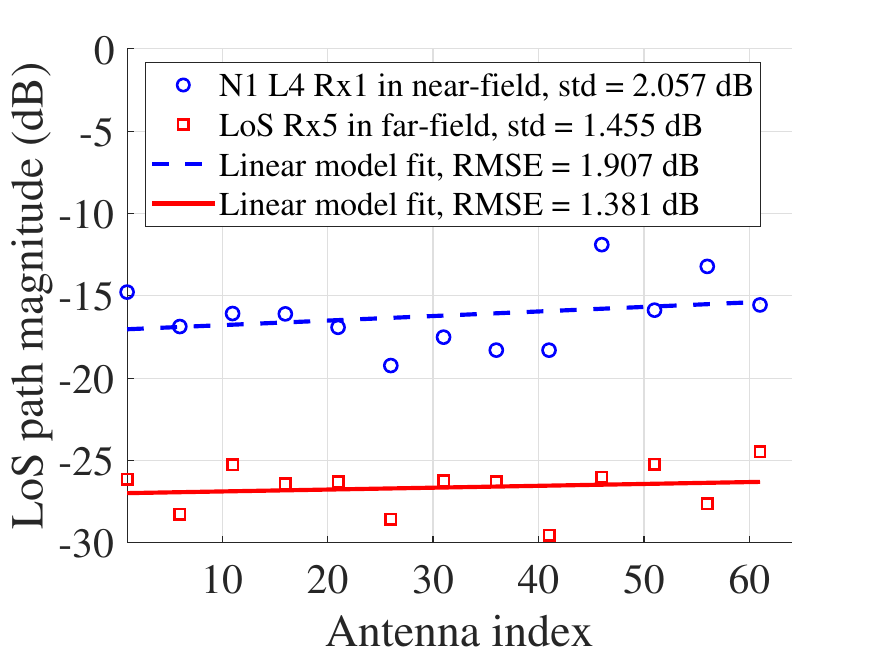}%
		\label{fig:magnitude}%
	}
	\subfloat[The LoS path phase.]{%
		\includegraphics[width=0.50\columnwidth]{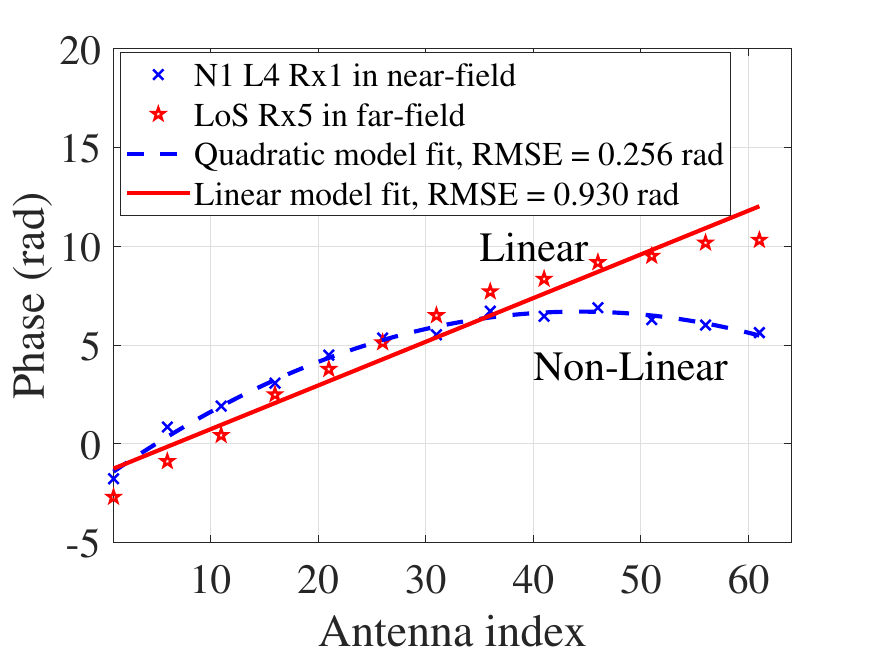}%
		\label{fig:phase}%
	}
	\caption{Near-field effects on (a) the LoS path magnitude and (b) the LoS path phase.} 
	\label{fig:Near-field effects}
\end{figure}
\section{New Channel Characterisation Analysis}\label{sec:new}
\subsection{Near-Field Effects}
As the array aperture increases, the near-field effects of UM-MIMO channels become increasingly pronounced, rendering the traditional far-field plane-wave assumption no longer valid. Current research on near-field effects in channel modeling has been conducted. For instance, simulations in \cite{NMB3} and \cite{NFeffects} have revealed non-linear phase phenomena in near-field spherical waves. However, near-field effects based on experimental measurements have not been fully verified or quantitatively analyzed. Consequently, this paper compares LoS channels under near-field and far-field conditions from the perspective of path power and phase evolution, with the aim of elucidating near-field propagation effects and providing quantitative support for the modeling of near-field LoS channels.

\subsubsection{Power}
In Fig. \ref{fig:Near-field effects}\subref{fig:magnitude}, compared with the far-field LoS channel, the LoS path power in the near-field LoS channel exhibits more pronounced spatial fluctuations across the array aperture. Specifically, in the near-field, the standard deviation of the LoS path power is 0.602 dB higher than that in the far-field, whilst the RMSE value of the near-field linear fit is 0.526 dB greater than that of the far-field linear fit. This indicates that under near-field propagation, the distribution of LoS energy received by each array element is no longer approximately uniform. Instead, it is jointly influenced by the spherical wavefront and by differences in the distance from the array element to the transmitter, resulting in greater power inconsistency within the aperture.
\subsubsection{Spherical-Wave Property}
In Fig. \ref{fig:Near-field effects}\subref{fig:phase}, the near-field effect is more pronounced in the phase dimension. The phase of the far-field LoS channel varies approximately linearly with the antenna index, with an RMSE of 0.930 rad for the linear fit, consistent with the propagation characteristics of plane waves. In contrast, the phase of the near-field LoS channel exhibits distinct non-linear curvature characteristics, with an RMSE of 0.256 rad when fitted using a quadratic model. This result indicates that near-field wavefront curvature cannot be ignored, and a non-linear phase model that accounts for spherical-wave propagation should therefore be adopted.

\subsection{Spatial Non-Stationarity}
SNS manifests itself through spatial variations in the statistical properties of the channel. A direct manifestation of the SNS is that the correlation between different antenna branches is not consistent across different locations. For MIMO channels, spatial correlation directly determines the degree of independence between antenna branches and consequently influences the spatial multiplexing and diversity performance of the system. Therefore, to quantitatively characterise the non-stationary features of the channel in the spatial dimension, the spatial cross-correlation function (SCCF) is employed as the metric. By normalising the inner product of the channel responses of different antennas, the SCCF effectively measures the degree of similarity between two antenna branches. A higher SCCF value indicates stronger channel correlation and lower independence, whereas a lower value indicates better channel independence. The SCCF \cite{RAA128x8PN}, \cite{VAAMulti-frequencyVNA}, \cite{SCCF} between different antennas at receivers and transmitters ${{l_{{\rm{Rx}}}}}$ and ${{l_{{\rm{Tx}}}}}$ respectively can be expressed as
\begin{equation}
	\mathop {SCCF}\limits_{{l_{{\rm{Rx}}}},{l_{{\rm{Tx}}}},q,p,t} = \frac{{\left| {{{\emph h}_{q,p}}\left( {{l_{{\rm{Rx}}}},{l_{{\rm{Tx}}}},t} \right) {\emph h}_{q',p}^{\bf{H}}\left( {{l_{{\rm{Rx}}}},{l_{{\rm{Tx}}}},t} \right)} \right|}}{{\Big\| {{\emph h}_{q,p}^{}\left( {{l_{{\rm{Rx}}}},{l_{{\rm{Tx}}}},t} \right)} \Big\| \cdot \Big\| { {\emph h}_{q',p}^{\bf{H}}\left( {{l_{{\rm{Rx}}}},{l_{{\rm{Tx}}}},t} \right)} \Big\|}}, \label{XX}
\end{equation}
where the superscript \textbf {H} indicates a conjugate transpose operation, the subscripts \emph {q} and \emph {\( q' \)} are different receiving antennas. $\left|  \cdot  \right|$ is the absolute value operator, $\left\|  \cdot  \right\|$ denotes the Euclidean norm of the vector. However, in \cite{RAA128x8PN}, \cite{VAAMulti-frequencyVNA}, \cite{SCCF}, the SCCF is essentially a pair-wise local correlation metric, and it cannot capture the overall correlation strength at different spatial locations or reveal the distribution patterns of channel SNS across different scenarios. Therefore, it is necessary to perform normalised cumulative processing on the SCCFs corresponding to all transmit-receive antenna combinations at a given location to construct the cumulative spatial cross-correlation function (CSCCF). The CSCCF is defined as
\begin{equation}	
\mathop {CSCCF}\limits_{{l_{{\rm{Rx}}}},{l_{{\rm{Tx}}}},t}  = \sum\limits_{q = 1}^{{N_{{\rm{Rx}}}}} {\sum\limits_{p = 1}^{{N_{Tx}}} {\frac{{SCCF\left( {{l_{{\rm{Rx}}}},{l_{{\rm{Tx}}}},q,p,t} \right)}}{{{N_{{\rm{Rx}}}}{N_{{\rm{Tx}}}}}}} }.   \label{XX}
\end{equation}

\begin{figure}[t]
	\centering
	\subfloat[]{%
		\includegraphics[width=0.7\columnwidth]{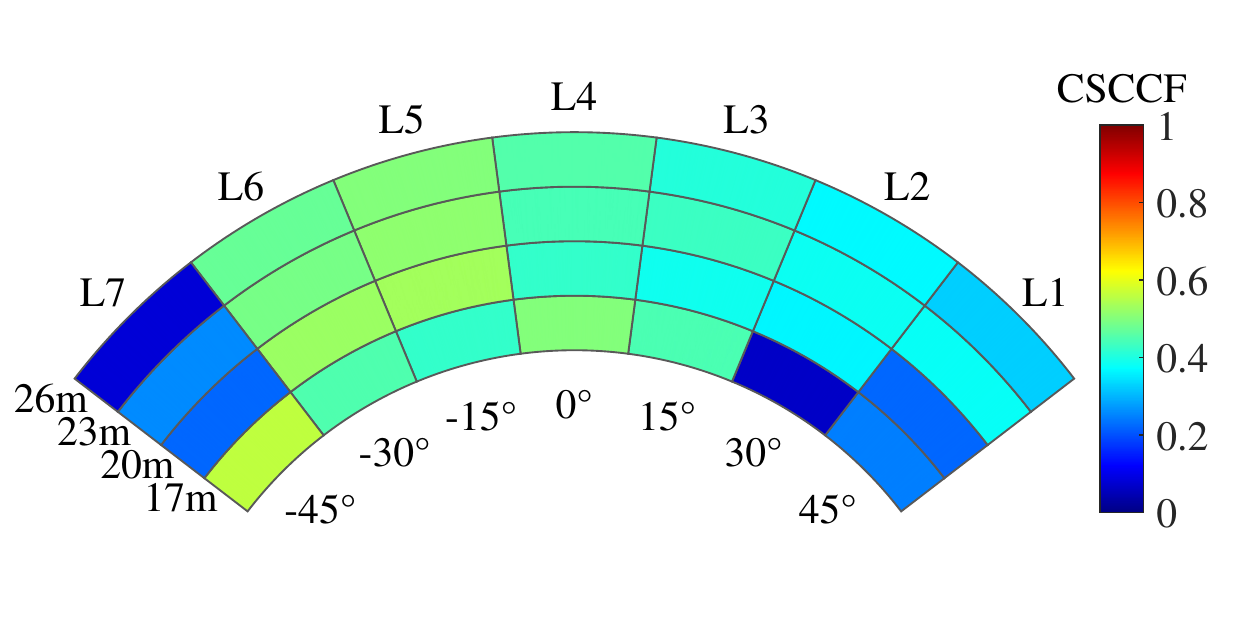}%
		\label{fig:N1 L1-L7}%
	}
	\vspace{-0.3cm}
	\subfloat[]{%
		\includegraphics[width=0.7\columnwidth]{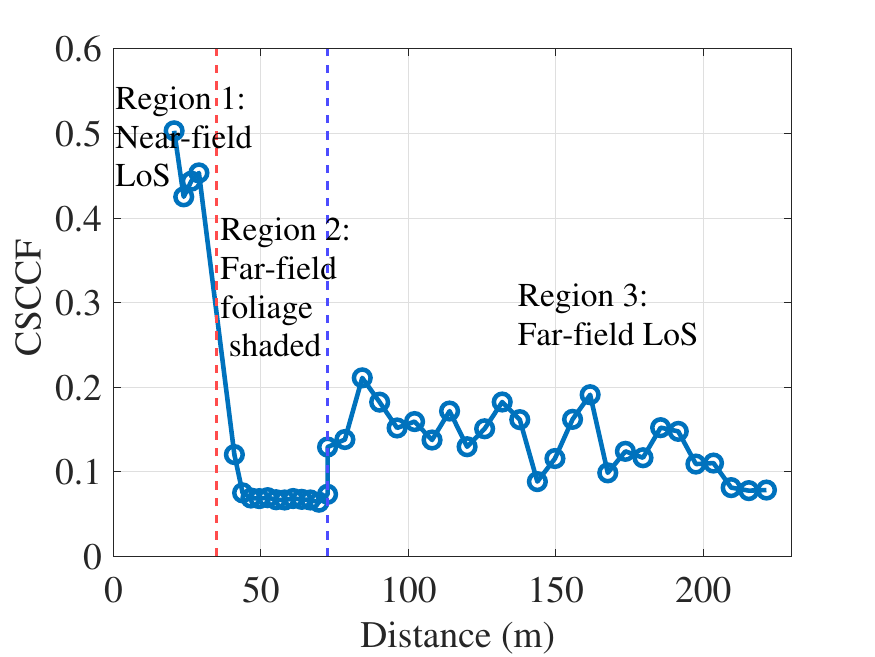}%
		\label{fig:NN to NF}%
	}
	\caption{The distribution of CSCCF values for (a) the near-field N1 L1-L7 and (b) the near-field L4 to the far-field foliage shaded route2, and then to the far-field LoS route3.} 
	\label{fig:CSSF}
\end{figure}

Fig. \ref{fig:CSSF} illustrates the distribution patterns of CSCCF values under different propagation conditions. Specifically, Fig. \ref{fig:CSSF}\subref{fig:N1 L1-L7} shows the variation in CSCCF values at the Rx points (Rx1--Rx4) along sub-routes L1 to L7 in the near-field N1 scenario. Under the near-field LoS conditions, the CSCCF values are relatively high, ranging from approximately 0.4 to 0.6, indicating strong signal correlation and poor channel independence. In particular, on the sub-route L4, the CSCCF values at the receiving points are close to 0.5, suggesting strong spatial correlation between signals. In contrast, in areas with leaf obstruction, the CSCCF values drop significantly to between 0.1 and 0.2, indicating that the spatial correlation of the signals is significantly weakened and channel independence is enhanced. Furthermore, Fig. \ref{fig:CSSF}\subref{fig:NN to NF} illustrates the variation in CSCCF values from the near-field LoS scenario to the far-field foliage shaded scenario, and then to the far-field LoS scenario. The CSCCF values at the near-field LoS propagation points (N1 L4 Rx1--Rx4) in region 1 are close to 0.5, indicating strong correlation between signals. Upon entering the region 2 (far-field leaf obstruction), the CSCCF values drop sharply to between 0.1 and 0.2, demonstrating that leaf obstruction effectively reduces the spatial correlation between signals, thereby enhancing channel independence. Finally, in the far-field LoS propagation in the region 3, the CSCCF values rise again to between 0.2 and 0.3. Although signal attenuation increases somewhat, the CSCCF values remain lower than those in the region 1 due to the greater distance of signal propagation and the lower spatial correlation.

Based on \cite{RAA128x8PN}, \cite{VAAMulti-frequencyVNA}, \cite{SCCF}, the CSCCF is further developed to provide a unified characterization of the overall spatial correlation of all transmit-receive antenna combinations at a given location. Using this metric, this paper further reveals the continuous evolution of spatial correlation between near-field LoS, far-field foliage shaded and far-field LoS conditions, and verifies the effectiveness of the CSCCF in characterising SNS and supporting subsequent channel modeling and MIMO system design.
\begin{figure}[t]
	\centering
	\includegraphics[width= 2.6 in]{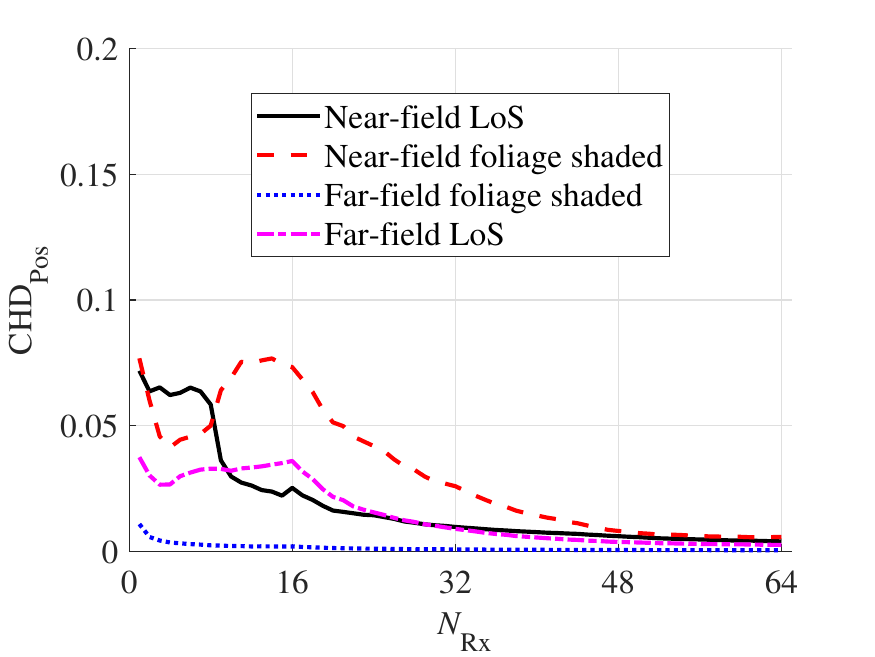}
	\caption{Channel hardening results for different scenarios when the number of transmit antennas is 128.}
	\label{fig:CHD}
\end{figure}
\subsection{Channel Hardening Characteristics}
As the size of the Rx array increases, the small-scale fading on individual antenna branches is progressively smoothed out during spatial summation, thereby reducing the instantaneous fluctuations in the equivalent channel gain and causing it to cluster more closely around its statistical mean. This phenomenon of fluctuation convergence resulting from large-scale arrays is known as channel hardening. The more pronounced the channel hardening, the lower the link’s sensitivity to random fading and the greater its transmission stability. To quantitatively characterise the degree of hardening under different scenarios, a spatial-domain channel hardening metric \cite{CHD1}, \cite{CHD2} based on measurement point statistics is employed, which can be expressed as

\begin{equation}
	CH{D_{{\rm{Pos}}}}\left( {{N_{{\rm{Rx}}}},f} \right) = \frac{{{\rm{Va}}{{\rm{r}}_m}\left\{ {\tilde g_m^{\left( {{N_{{\rm{Rx}}}}} \right)}\left( f \right)} \right\}}}{{{{\left( {\mathbb{E}{_m}\left\{ {\tilde g_m^{\left( {{N_{{\rm{Rx}}}}} \right)}\left( f \right)} \right\}} \right)}^2}}},\label{CHD_Pos}
\end{equation}
where ${\rm{Va}}{r_m}\left\{  \cdot  \right\}$ and ${\mathbb{E}_m}\left\{  \cdot  \right\}$ denote the variance and expectation with respect to the point index $m \in \left\{ {1, \ldots ,M} \right\}$ , respectively. $\tilde g_m^{\left( {{N_{{\rm{Rx}}}}} \right)}\left( f \right)$ represents the normalized synthetic gain at frequency point $f$ for point $m$, synthesized by ${N_{{\rm{Rx}}}}$ receiving elements. This can be defined as
\begin{equation}
	\tilde g_m^{\left( {{N_{{\rm{Rx}}}}} \right)}\left( f \right) \buildrel \Delta \over = \left\| {{\bf{\tilde H}}_m^{\left( {{N_{{\rm{Rx}}}}} \right)}\left( f \right)} \right\|_F^2,\label{}
\end{equation}
where ${\left\|  \cdot  \right\|_F}$ denotes the Frobenius norm. ${\bf{\tilde H}}_m^{\left( {{N_{{\rm{Rx}}}}} \right)}\left( f \right)$ is defined as the normalized subchannel frequency-domain matrix at point $m$, which can be obtained by selecting the array elements according to the normalized frequency domain matrix ${{\bf{\tilde H}}_m}\left( f \right)$. ${{\bf{\tilde H}}_m}\left( f \right)$ can be defined as
\begin{equation}
	{{\bf{\tilde H}}_m}\left( f \right) \buildrel \Delta \over = {{{{\bf{H}}_m}\left( f \right)} \mathord{\left/
			{\vphantom {{{{\bf{H}}_m}\left( f \right)} {\sqrt {{\alpha _m}} }}} \right.
			\kern-\nulldelimiterspace} {\sqrt {{\alpha _m}} }}, \label{}
\end{equation}
where ${{\bf{H}}_m}\left( f \right)$ is  the MIMO frequency-domain channel matrix measured at point $m$. ${\alpha _m}$ is the energy normalization coefficient for point $m$, which can be defined as 
\begin{equation}
	{\alpha _m} \buildrel \Delta \over = {\mathbb{E}_f}\left\{ {\left\| {{{\bf{H}}_m}\left( f \right)} \right\|_F^2} \right\}, \label{}
\end{equation}
where ${\mathbb{E}_f}\left\{  \cdot  \right\}$ indicates the averaging operation across each frequency point within the measurement bandwidth. $\left\| {{{\bf{H}}_m}\left( f \right)} \right\|_F^2$ denotes the total energy of the channel matrix at frequency point $f$ for point $m$.
To derive a scalar metric that facilitates cross-comparisons between different array sizes and measurement scenarios, we aggregate the results in the frequency domain based on Equation (21) to obtain the channel hardening criterion:
\begin{equation}
	CH{D_{{\rm{Pos}}}}\left( {{N_{{\rm{Rx}}}}} \right) = {\rm{media}}{{\rm{n}}_f}\left\{ {CH{D_{{\rm{Pos}}}}\left( {{N_{{\rm{Rx}}}},f} \right)} \right\}, \label{}
\end{equation}
where ${\rm{media}}{{\rm{n}}_f}\left\{  \cdot  \right\}$ denotes the median value across all frequency points within the entire measurement bandwidth.

Fig. \ref{fig:CHD} shows the variation in CHD values with respect to ${N_{{\rm{Rx}}}}$ across different scenarios under the condition of ${N_{{\rm{Tx}}}}$ = 128. Overall, as the number of receiving elements increases, the CHD values in all four scenarios ultimately exhibit a downward trend, indicating that increasing the size of the receiving array can effectively mitigate fluctuations in equivalent channel gain between different measurement points, thereby enhancing the channel hardening effect. This is consistent with the general trend reported in \cite{RAA128x8PN}. Across the different scenarios, the far-field foliage shaded scenario consistently exhibits the lowest CHD values, demonstrating the strongest channel-hardening effect. The CHD values in the near-field foliage shaded scenario are generally higher and exhibit local peaks at ${N_{{\rm{Rx}}}}$ = 8--16, indicating stronger channel fluctuations and a need for larger array sizes to achieve stable hardening gains. In contrast, the two LoS scenarios exhibit certain differences under small-scale arrays, but gradually converge as ${N_{{\rm{Rx}}}}$ increases, suggesting that spatial averaging effects dominate under large-scale arrays, whilst the marginal benefits of adding further elements gradually diminish.

In summary, the results confirm the general conclusion in \cite{RAA128x8PN}, \cite{CHD3} that larger arrays improve channel hardening, while further showing that this effect is highly scenario-dependent and influenced by propagation conditions. Based on measurement results, this work provides practical insight into the evolution of channel hardening under different deployment scenarios, and thus offers useful guidance for UM-MIMO array design and link reliability evaluation in realistic environments.

\section{UM-MIMO Communication System Performance Analysis}\label{sec:Systems}

\subsection{System Channel Capacity}
\par When data signals are transmitted over the measured channel, the channel capacity \emph{C} \cite{Capacity1}, \cite{Capacity3} of the system is given by
\begin{equation}
	\begin{split}
		C = &\frac{1}{{{N_s}{N_f}}}\sum\limits_{j = 1}^{{N_s}} {\sum\limits_{k = 1}^{{N_f}} {{{\log }_2}} }
		\det \Big( {{{\bf{I}}_{{N_{{\rm{Rx}}}}}}}+ 	\\
		& \frac{\rho }{{{N_{{\rm{Tx}}}}}}{\bf{\mathord{\buildrel{\lower3pt\hbox{$\scriptscriptstyle\frown$}}
					\over H} }}\left( {j,k} \right){{{\bf{\mathord{\buildrel{\lower3pt\hbox{$\scriptscriptstyle\frown$}}
							\over H} }}}^{\rm{H}}}\left( {j,k} \right) \Big),
	\end{split}
\end{equation}
where $\det \left(  \cdot  \right)$ is the determinant of the matrix, ${{\mathbf{I}}_{{{N}_{\text{Rx}}}}}$ is the ${{N}_{\text{Rx}}}\times {{N}_{\text{Rx}}}$ unit matrix, $\rho$ represents the signal to noise ratio (SNR). ${\bf{\mathord{\buildrel{\lower3pt\hbox{$\scriptscriptstyle\frown$}} \over H} }}\left( {j,k} \right)$ represents the normalized channel matrix at the $j$-th snapshot and $k$-th frequency point, and can be expressed as ${\bf{\mathord{\buildrel{\lower3pt\hbox{$\scriptscriptstyle\frown$}}\over H} }}\left( {j,k} \right) = {{{\bf{H}}\left( {j,k} \right)} \mathord{\left/	{\vphantom {{{\bf{H}}\left( {j,k} \right)} {\sqrt {{P_{rs}}\left( {j,k} \right)} }}} \right. \kern-\nulldelimiterspace} {\sqrt {{P_{\rm rs}}\left( {j,k} \right)} }}$. Here, ${P_{\rm rs}}\left( {j,k} \right) = \frac{1}{{{N_{{\rm{Rx}}}}{N_{{\rm{Tx}}}}}}\left\| {{\bf{H}}\left( {j,k} \right)} \right\|_{\rm{F}}^2$.
\begin{figure}[t]
	\centering
	\includegraphics[width= 2.6 in]{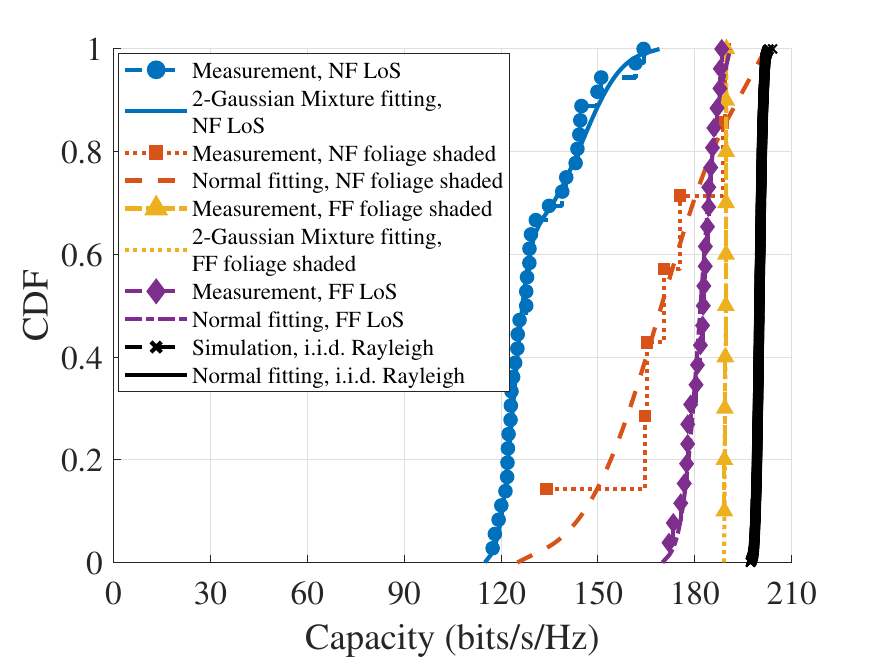}
	\caption{The CDFs of channel capacity for different scenarios when $\rho = 10$ dB.}
	\label{fig:Capacity}
\end{figure}
\vspace{0.2cm}
\par Fig. \ref{fig:Capacity} shows the CDFs of channel capacity for different scenarios at $\rho $ = 10 dB. As shown in Fig. \ref{fig:Capacity}, the CDFs of channel capacity can be fitted by two-Gaussian mixture distribution with $0.367{\cal N}\left( {145,{{9.8}^2}} \right) + 0.633{\cal N}\left( {124,{{3.56}^2}} \right)$ and $0.405{\cal N}\left( {189,{{0.123}^2}} \right) + 0.595{\cal N}\left( {190,{{0.055}^2}} \right)$ for NF LoS and FF foliage shaded scenarios, respectively. The CDFs of channel capacity can be fitted by normal distribution with ${\cal N}\left( {169.7,{{18.68}^2}} \right)$ and ${\cal N}\left( {181.8,{{4.47}^2}} \right)$ for NF foliage shaded and FF LoS scenarios, respectively. Furthermore, the CDF fitting distribution of the i.i.d. Rayleigh MIMO channel reference model with the identical array size and the SNR follows a normal distribution ${\cal N}\left( {200,{{0.987}^2}} \right)$.

From the fitting results, it can be seen that when $\rho $ = 10 dB, the average channel capacities for the NF LoS, NF foliage-shaded, FF LoS and FF foliage-shaded scenarios are 131.7 bits/s/Hz, 169.7 bits/s/Hz, 181.8 bits/s/Hz and 189.6 bits/s/Hz respectively, all of which are lower than the 200.0 bits/s/Hz obtained for the i.i.d. Rayleigh reference channel, with corresponding capacity differences of 68.3 bits/s/Hz, 30.3 bits/s/Hz, 18.2 bits/s/Hz and 10.4 bits/s/Hz. It can be seen that NF LoS deviates most significantly from the reference channel, indicating that when the dominant LoS path is strong in the near field, channel correlation is stronger and spatial freedom is more significantly constrained. Meanwhile, compared to the LoS scenario, the average capacity in the foliage-shaded scenario increases by 38.0 bits/s/Hz in the near field and 7.8 bits/s/Hz in the far field, indicating that the additional scattering introduced by the vegetation obstruction helps to enhance the channel rank and spatial multiplexing capability. Meanwhile, compared to the NF LoS and NF foliage-shaded scenarios, the FF LoS and FF foliage-shaded scenarios exhibit average channel capacity increases of 50.1 bits/s/Hz and 19.9 bits/s/Hz, respectively. The average channel capacity generally follows the order `FF foliage-shaded $\textgreater$ FF LoS $\textgreater$ NF foliage-shaded $\textgreater$ NF LoS', indicating that channel capacity under far-field and high-scattering conditions is closer to that of the Rayleigh reference channel.

Based on actual channel measurements, this paper provides a statistical characterization and analysis of MIMO channel capacity distributions under near-field LoS, far-field LoS, near-field foliage shaded and far-field foliage shaded conditions in the UMi environment, which will provide a crucial basis for the evaluation of UM-MIMO systems.

\section{Conclusion}\label{sec:Conclusion}
In this paper, UM-MIMO channel measurement campaigns were conducted in the UMi environment at 15 GHz. Four representative propagation scenarios were investigated, including near-field LoS, near-field foliage-shaded, far-field foliage-shaded, and far-field LoS. A 64 $\times$ 128 UM-MIMO channel sounder was developed based on a L-shaped antenna array at Tx and a square planar array at Rx. The results of this work provide measurement-based insights into the propagation characteristics of UM-MIMO channels in 15 GHz UMi environments. The extracted path loss and foliage attenuation parameters can serve as useful references for channel modeling and propagation prediction in similar scenarios. Furthermore, the analysis of near-field effects, spatial non-stationarity, and channel hardening improves the understanding of the unique propagation behavior of UM-MIMO systems. Overall, this study provides both theoretical support and practical guidance for the modeling, design, and performance evaluation of future UM-MIMO wireless communication systems.

\begingroup
\normalsize          
\bibliographystyle{IEEEtran}
\bibliography{reference}
\endgroup

\end{document}